\documentclass[numberedappendix]{emulateapj}

\usepackage{mathrsfs}
\usepackage{color}
\newcommand{\vc}[1]{\mbox{\boldmath{$#1$}}}
\newcommand{\de}{\mathrm{d}}
\newcommand{\dpa}{\partial}

\newcommand{\nab}{\vc{\nabla}}

\newcommand{\ii}{\mathrm i}
\DeclareMathSymbol{\varOmega}{\mathord}{letters}{"0A}
\DeclareMathSymbol{\varSigma}{\mathord}{letters}{"06}
\DeclareMathSymbol{\varPsi}{\mathord}{letters}{"09}
\DeclareMathSymbol{\varPhi}{\mathord}{letters}{"08}
\DeclareMathSymbol{\varGamma}{\mathord}{letters}{"00}

\newcommand{\Eq}[1]{equation (\ref{#1})}
\newcommand{\Eqs}[2]{equations (\ref{#1}) and~(\ref{#2})}

\newcommand{\App}[1]{Appendix~\ref{#1}}

\newcommand{\Fig}[1]{Fig.~\ref{#1}}

\newcommand{\Tab}[1]{Table \ref{#1}}

\newcommand{\bl}{{\textcolor{white}{1}}}

\shorttitle{Zonal Flows in Magnetorotational Turbulence}
\shortauthors{Johansen, Youdin, \& Klahr}

\begin{document}

\title{Zonal Flows and Long-Lived Axisymmetric Pressure Bumps\\in
Magnetorotational Turbulence}

\author{A. Johansen}
\affil{Leiden Observatory, Leiden University, P.O.\ Box 9513, 2300 RA Leiden, The Netherlands}
\email{ajohan@strw.leidenuniv.nl}

\author{A. Youdin}
\affil{Canadian Institute for Theoretical Astrophysics, University of Toronto,
60 St. George Street, Toronto, Ontario M5S 3H8, Canada}
\email{youd@cita.utoronto.ca}

\and

\author{H. Klahr}
\affil{Max-Planck-Institut f\"ur Astronomie, 69117 Heidelberg, Germany}
\email{klahr@mpia.de}

\begin{abstract}
We study the behavior of magnetorotational turbulence in shearing box
simulations with a radial and azimuthal extent up to ten scale heights. Maxwell
and Reynolds stresses are found to increase by more than a factor two when
increasing the box size beyond two scale heights in the radial direction.
Further increase of the box size has little or no effect on the statistical
properties of the turbulence. An inverse cascade excites magnetic field
structures at the largest scales of the box. The corresponding 10\% variation
in the Maxwell stress launches a zonal flow of alternating sub- and
super-Keplerian velocity. This in turn generates a banded density structure in
geostrophic balance between pressure and Coriolis forces. We present a
simplified model for the appearance of zonal flows, in which stochastic forcing
by the magnetic tension on short time-scales creates zonal flow structures with
life-times of several tens of orbits. We experiment with various improved
shearing box algorithms to reduce the numerical diffusivity introduced by the
supersonic shear flow. While a standard finite difference advection scheme
shows signs of a suppression of turbulent activity near the edges of the box,
this problem is eliminated by a new method where the Keplerian shear advection
is advanced in time by interpolation in Fourier space.
\end{abstract}
\keywords{diffusion --- hydrodynamics --- instabilities --- planetary systems:
protoplanetary disks --- solar system: formation --- turbulence}

\section{Introduction}

Turbulence forms a main pillar of modern accretion disk theory \citep[see
review by][]{BalbusHawley1998}. The most promising turbulence candidate is the
magnetorotational instability \citep[MRI, ][]{BalbusHawley1991}, which renders
Keplerian rotation profiles linearly unstable in the presence of a magnetic
field of moderate strength. The non-linear evolution of the MRI is often
modeled in a local, corotating box representing a small section of a Keplerian
disk \citep[e.g.][]{Hawley+etal1995,Brandenburg+etal1995}. This allows high
resolution studies with periodic boundary conditions. However, most shearing
box simulations have been limited to a narrow radial extent of approximately
one scale height.

In this paper we explore shearing box simulations with a radial domain up to
ten scale heights. Increasing the radial extent helps bridge the gap between
smaller local simulations and global simulations of accretion disks
\citep[e.g.][]{Armitage1998,ArltRuediger2001,FromangNelson2006,Lyra+etal2008}.
Since rotating shear flows are prone to an inverse cascade of magnetic energy
\citep{BrandenburgSubramanian2005}, large simulation domains are needed to
fully capture the saturated state of the MRI.

We find that the statistical properties of magnetorotational turbulence
converge relatively well as we increase the radial extent of the simulation box
beyond a couple of scale heights. However, large shearing boxes do not behave
simply like multiple copies of moderately sized boxes. Axisymmetric pressure
bumps, in geostrophic balance with a sub-Keplerian/super-Keplerian zonal flow
envelope, grow to fill the radial width of the box. A main purpose of this
paper is to describe the excitation and dynamics of such zonal flows.

The emergence of zonal flows in protoplanetary disks can have a profound
influence on the growth of terrestrial planets and giant planet cores. Planet
formation begins as dust grains collide and grow to ever larger bodies
\citep{Safronov1969,Dominik+etal2007,BlumWurm2008}. Macroscopic solids drift
rapidly through the disk, due to the head wind from the slightly sub-Keplerian
gas \citep{Weidenschilling1977}, imposing a severe time-scale constraint on the
growth from approximately cm-sized pebbles to bodies several kilometers in
size. A weak variation of the rotation profile caused by zonal flows may
nevertheless suffice to locally stop the radial drift flow \citep{Whipple1972,
KlahrLin2001,HaghighipourBoss2003,FromangNelson2005,Kato+etal2008}. Such a
convergence zone can quickly accumulate rocks and boulders, enhancing the local
density enough to trigger gravitational instabilities which form planetesimals
or dwarf planets in a very short time
\citep{YoudinShu2002,Johansen+etal2006,Johansen+etal2007}. The reduced radial
drift due to persistent zonal flows may also facilitate dust growth by a longer
time-scale coagulation-fragmentation process
\citep{Weidenschilling1997,DullemondDominik2005,Brauer+etal2008a,Brauer+etal2008b,Johansen+etal2008}.

Turbulent density fluctuations exert significant gravitational torques on
planetesimals and planetary embryos. This adds a stochastic element to the
normally inward type-I planetary migration from gravitational interactions with
the gas disk. Migration can be slowed, sped up, or even reversed, especially
for lower mass planets
\citep{NelsonPapaloizou2004,Laughlin+etal2004,Johnson+etal2006,Oishi+etal2007}.
Gravitational scattering can also pump planetesimal eccentricities to the level
where collisions become destructive for small planetesimals with low surface
gravity \citep{Ida+etal2008}. The dominant \emph{axisymmetric} component of the
density fluctuations that we analyze in this paper exerts no gravitational
torques. Thus non-axisymmetric modes -- perhaps in global simulations -- must
be considered in a future publication for applications to planet migration and
scattering.

Zonal flows arise in a diverse range of physical settings. The launching
mechanism in the examples below differ, both from each other and from the
problem at hand. A common feature is non-linear mode coupling and variation in
turbulent transport coefficients over large length scales.

{\it Giant planet atmospheres}. The banded structure in Jupiter's cloud layer
is perhaps the most famous manifestation of zonal flows. \cite{Busse1976}
proposed that these flows arise from an inverse cascade of thermally driven
convection in giant planet interiors. Simulations of convective turbulence in
spherical shells \citep[e.g.,][]{Sun+etal1993,HeimpelAurnou2007} reproduce the
zonal flows in gas and ice giants both qualitatively and quantitatively.

{\it Torsional oscillations}. On top of the differential rotation profile of
the Sun there is a zonal flow (with 3 m/s amplitude) that migrates from high to
low latitudes during the solar cycle \citep{HowardLabonte1980}. The zonal flow
is attributed to the toroidal component of the Lorentz force exerted by
magnetic fields arising in the solar dynamo
\citep{Schuessler1981,Yoshimura1981}.

{\it Laboratory plasmas}. Laboratory plasmas often exhibit strong zonal flows.
These self-organized structures arise from a non-local inverse cascade mediated
by drift waves. The excellent review by \cite{Diamond+etal2005} points out the
intimate mathematical equivalence between Rossby wave dynamics in planetary
atmospheres and drift wave dynamics in plasmas.

In magnetorotational accretion disk turbulence we find that the zonal flow is
excited by a large scale variation in the Maxwell stress. The differential
momentum transport by the associated magnetic tension launches the
sub-Keplerian/super-Keplerian zonal flows, which in turn are slightly
compressive and form pressure bumps with life-times approximately equal to the
turbulent mixing time-scale. The variation in the Maxwell stress is a
consequence of an inverse cascade of magnetic energy to the largest scales of
the box.

Practical numerical issues arise in simulations of large shearing boxes, since
the supersonic orbital shear introduces significant numerical diffusivity. In
simulations of magnetorotational turbulence, \cite{Johnson+etal2008} found a
spurious density depression in the radial center of the grid. Here the velocity
of the shear flow vanishes, minimizing numerical dissipation, and the local
increase in turbulent stresses leads to a mass transport away from the box
center. We also find indications of this unphysical behavior in our biggest
simulations, if we use the standard shearing box solver of the Pencil Code.
However we can essentially eliminate this undesired effect with two independent
techniques described in the appendices: (1) separately solving the known
advection by orbital shear with high order interpolation or (2) displacing the
entire grid systematically in the radial direction so that preferred locations
in the box vary too rapidly to generate artificial structures. This gives us
confidence that the zonal flows we see are not a numerical artifact.

The structure of the paper is as follows. In \S\ref{s:setup} we describe the
shearing box coordinate frame, dissipation operators and simulation parameters.
The properties of turbulence in large shearing boxes are presented in
\S\ref{s:turbulence}. The following section, \S\ref{s:zonalflows}, is dedicated
to describing the large scale zonal flows that are ubiquitous to the
simulations. In \S\ref{s:zonalflowmodel} we present a simplified model for the
stochastic excitation of zonal flows. The radial variation in Maxwell stress
and the inverse cascade of magnetic energy is discussed in detail in
\S\ref{s:cascade}. In \S\ref{s:validation} we validate our results by
presenting convergence tests, variations of the main simulations using
different shearing box algorithms, and the dependence of the zonal flows on the
order and magnitude of the explicit dissipation. In \S\ref{s:discussion} we
summarize our results briefly and discuss caveats and implications. The
appendices contain a test of the shearing box solver (\App{s:shearwave}), a
description of a newly developed technique to interpolate Keplerian advection
in wavenumber space (\App{s:safi}), and additional techniques to reduce
space-dependent numerical diffusivity (\App{s:xdepdiff}).

\section{Model Setup}
\label{s:setup}

We simulate a local, corotating patch of a Keplerian disk in the shearing box
approximation. The axes are oriented such that $x$ points outwards along the
cylindrical radius, $y$ points along the main rotation direction, and $z$
points vertically out of the disk parallel to the Keplerian rotation vector
$\vc{\varOmega}$. The criterion for the validity of the shearing box
approximation is strictly that the box size $L$ should be much smaller than the
orbital radius $r$, so that curvature terms may be ignored. In practice this
implies that the box size measured in scale heights, $L/H$, must be much
smaller than the inverse aspect ratio of the disk, $(L/H)\ll 1/(H/r)$. For
typical values $(H/r)=0.01\ldots0.1$ we get $(L/H)\ll10\ldots100$. We shall
consider boxes with a radial extent of up to ten times the scale height, which
is still fairly well represented in the shearing box approximation. We have
chosen the shearing box approximation over global disk simulations because of
the absence of boundaries and because it makes it possible to study convergence
in resolution and box size from a relatively well understood coordinate frame.

The ideal MHD equation of motion, for the velocity field $\vc{u}$ relative to
the linearized Keplerian shear, is
\begin{eqnarray}\label{eq:eqmot}
  \frac{\dpa \vc{u}}{\dpa t} +
    (\vc{u} \cdot \nab) \vc{u} + u_y^{(0)} \frac{\dpa \vc{u}}{\dpa y} &=& 
    2 \varOmega u_y \vc{e}_x - \frac{1}{2} \varOmega u_x
    \vc{e}_y - \varOmega^2 z \vc{e}_z \nonumber \\
    & & \hspace{-1.0cm} + \frac{1}{\rho} \vc{J} \times \vc{B} -
    \frac{1}{\rho} \nab P + \vc{f}_\nu(\vc{u},\rho) \, .
\end{eqnarray}
The left hand side of the equation contains the advection by the perturbed
velocity $\vc{u}$ and by the Keplerian shear flow $u_y^{(0)}\vc{e}_y$, where
$u_y^{(0)}=-(3/2)\varOmega x$. The right hand side contains the usual terms for
Coriolis force, vertical component of the gravity of the central object,
Lorentz force, pressure gradient force and an explicit viscosity term
$\vc{f}_\nu$ constructed to dissipate the kinetic energy released by Reynolds
and Maxwell stresses (see \S\ref{s:dissipation} for details). The magnetic
field is calculated from the magnetic vector potential through
$\vc{B}=\nab\times\vc{A}$, and the current density is calculated through
Ampere's law $\vc{J}=\mu_0^{-1} \nab\times(\nab\times\vc{A})$.

The magnetic vector potential $\vc{A}$ is evolved through the uncurled
induction equation
\begin{equation}\label{eq:ind}
  \frac{\dpa \vc{A}}{\dpa t} + u_y^{(0)} \frac{\dpa \vc{A}}{\dpa y} = \vc{u}
  \times \vc{B} +\frac{3}{2} \varOmega A_y \vc{e}_x + \vc{f}_\eta(\vc{A}) 
  \, .
\end{equation}
The terms on the right hand side are the electromotive force and an explicit
magnetic stretching term representing the creation of $A_x$ from $A_y$
(conversely creation of $B_y$ from $B_x$) by the Keplerian shear. Explicit
resistivity is taken into account through the term $\vc{f}_\eta$ (see
\S\ref{s:dissipation}). For simulations with a weak imposed vertical field, we
add a constant field $\langle B_z\rangle\vc{e}_z$ to the magnetic field
$\vc{B}$ appearing in the electromotive force in \Eq{eq:ind} and in the Lorentz
force in \Eq{eq:eqmot}. In the analysis of our results we will seek to
understand the evolution of the magnetic field via the action of individual
terms from the induction equation for $\vc{B}$, so we give it here for
reference,
\begin{equation}\label{eq:eqindb}
  \frac{\dpa \vc{B}}{\dpa t} + (\vc{u}\cdot\nab)\vc{B} + u_y^{(0)} \frac{\dpa
  \vc{B}}{\dpa y} =
  (\vc{B}\cdot\nab)\vc{u}-\frac{3}{2}\varOmega B_x \vc{e}_y
  -\vc{B}\nab\cdot\vc{u} \, .
\end{equation}
The advection terms appear in their usual form on the left hand side. The terms
on the right hand side of \Eq{eq:eqindb} are the internal stretching term, the
Keplerian stretching term, and the compression term.

Finally the mass density $\rho$ is evolved through the continuity equation
\begin{equation}\label{eq:cont}
  \frac{\dpa \rho}{\dpa t} + \vc{u}\cdot\nab\rho +
  u_y^{(0)}\frac{\dpa \rho}{\dpa y} = - \rho \nab
  \cdot \vc{u} + f_{\rm D}(\rho) \, .
\end{equation}
The last term on the right hand side is an explicit mass diffusion term
(presented in \S\ref{s:dissipation}). We use an isothermal equation of state,
$P=c_{\rm s}^2\rho$, to connect the pressure $P$ and the density $\rho$. The
soundspeed $c_{\rm s}$ is assumed to be constant.

\begin{deluxetable*}{lccccccccr}
  \tabletypesize{\tiny}
  \tablecaption{Run Parameters}
  \tablewidth{0pt}
  \tablehead{
    \colhead{Run} &
    \colhead{$L_x\times L_y\times L_z$} &
    \colhead{$N_x\times N_y\times N_z$} &
    \colhead{$\nu_i$} &
    \colhead{$\eta_i$} &
    \colhead{Order} &
    \colhead{Strat.} &
    \colhead{$\langle B_z \rangle$} &
    \colhead{Shear} &
    \colhead{$\Delta t$} }
  \startdata
    S          & $ 1.32\times 1.32\times 5.28$ & $\bl32\times\bl32\times128$ &
                 $6.0\times10^{-10}$ & $6.0\times10^{-10}$ & $3$ &
                 Yes & $0.0$ & FDA & $100$ \\
    M          & $ 2.64\times 2.64\times 5.28$ & $\bl64\times\bl64\times128$ &
                 $6.0\times10^{-10}$ & $6.0\times10^{-10}$ & $3$ &
                 Yes & $0.0$ & FDA & $100$ \\
    L          & $ 5.28\times 5.28\times 5.28$ & $128\times128\times128$ &
                 $6.0\times10^{-10}$ & $6.0\times10^{-10}$ & $3$ &
                 Yes & $0.0$ & SAFI & $100$ \\
    H          & $10.56\times10.56\times 5.28$ & $256\times256\times128$ &
                 $6.0\times10^{-10}$ & $6.0\times10^{-10}$ & $3$ &
                 Yes & $0.0$ & SAFI & $100$ \\
    M\_r2      & $ 2.64\times 2.64\times 5.28$ & $128\times128\times256$ &
                 $2.0\times10^{-11}$ & $2.0\times10^{-11}$ & $3$ &
                 Yes & $0.0$ & FDA & $100$ \\
    L\_nogz    & $ 5.28\times 5.28\times 5.28$ & $128\times128\times128$ &
                 $6.0\times10^{-10}$ & $6.0\times10^{-10}$ & $3$ &
                 No  & $0.0$ & SAFI & $100$ \\
    \hline
    M\_Bz\_r0.5& $ 2.64\times 2.64\times 5.28$ & $\bl32\times\bl32\times\bl64$ &
                 $2.0\times10^{-8\bl}$ & $2.0\times10^{-8\bl}$ & $3$ &
                 Yes & $0.01$ & FDA & $100$ \\
    M\_Bz      & $ 2.64\times 2.64\times 5.28$ & $\bl64\times\bl64\times128$ &
                 $6.0\times10^{-10}$ & $6.0\times10^{-10}$ & $3$ &
                 Yes & $0.01$ & FDA & $100$ \\
    M\_Bz\_r2  & $ 2.64\times 2.64\times 5.28$ & $128\times128\times256$ &
                 $2.0\times10^{-11}$ & $2.0\times10^{-11}$ & $3$ &
                 Yes & $0.01$ & FDA &  $50$ \\
    \hline
    MS         & $ 2.64\times 2.64\times 1.32$ & $\bl64\times\bl64\times\bl32$ &
                 $6.0\times10^{-10}$ & $6.0\times10^{-10}$ & $3$ &
                 No & $0.0$ & SRD & $100$ \\
    MS\_r2     & $ 2.64\times 2.64\times 1.32$ & $128\times128\times\bl64$ &
                 $2.0\times10^{-11}$ & $2.0\times10^{-11}$ & $3$ &
                 No & $0.0$ & FDA & $100$ \\
    MS\_r2\_fix& $ 2.64\times 2.64\times 1.32$ & $128\times128\times\bl64$ &
                 $6.0\times10^{-10}$ & $6.0\times10^{-10}$ & $3$ &
                 No & $0.0$ & SRD & $100$ \\
    \hline
    SS\_r4\_nu1& $ 1.32\times 1.32\times 1.32$ & $128\times128\times128$ &
                 $3.0\times10^{-4\bl}$ & $8.0\times10^{-5\bl}$ & $1$ &
                 No & $0.0$ & FDA & $40$ \\
    SS\_r8\_nu1& $ 1.32\times 1.32\times 1.32$ & $256\times256\times256$ &
                 $3.0\times10^{-4\bl}$ & $8.0\times10^{-5\bl}$ & $1$ &
                 No & $0.0$ & FDA & $25$ \\
    MS\_r4\_nu1& $ 2.64\times 2.64\times 1.32$ & $256\times256\times128$ &
                 $3.0\times10^{-4\bl}$ & $8.0\times10^{-5\bl}$ & $1$ &
                 No & $0.0$ & FDA & $50$ \\
  \enddata
  \tablecomments{Col.\ (1): Name of run. Col.\ (2): Box size in units of scale
  heights. Col.\ (3): Grid resolution. Col.\ (4): Viscosity coefficient. Col.\
  (5): Resistivity coefficient. Col.\ (6): Dissipation order. Col.\ (7):
  Stratification. Col.\ (8): Mean imposed vertical field. Col.\ (9): Shear
  advection scheme. Col.\ (10): Total run time in orbits ($T_{\rm orb} =
  2\pi\varOmega^{-1})$.}
  \label{t:runs}
\end{deluxetable*}
As a numerical solver we use the 6th order symmetric finite difference code
Pencil Code\footnote{The code is accessible for download at\\
\url{http://www.nordita.org/software/pencil-code/};\\ see
\cite{Brandenburg2003} for details on the numerical algorithm of the Pencil
Code.}. In Appendix \ref{s:shearwave} we test the standard shearing box
algorithm of the Pencil Code against a time-dependent analytical solution
\citep{BalbusHawley2006} for a shearing wave and find excellent agreement. We
also compare the evolution of a low amplitude, magnetized shear wave to a
numerical integration of the linearized evolution equations. However, while the
tests in Appendix \ref{s:shearwave} give confidence that the Pencil Code solves
the shearing box equations correctly, they are not sufficient to fully validate
our non-linear simulations. For this we will present the main results using
several independent shearing box algorithms. Normally we use the same finite
difference advection scheme for the Keplerian shear advection as for any other
advection term, but we have also implemented the Keplerian advection as an
interpolated shift of the dynamical variables \citep[similar
to][]{Masset2000,Gammie2001,Johnson+etal2008}. This has the advantage that the
space-dependence of the numerical diffusivity is removed and that the time-step
constraint connected with the Keplerian shear flow is eased. Our new Shear
Advection by Fourier Interpolation (SAFI) scheme, described in detail in
\App{s:safi}, differs from other shear advection schemes in that interpolation
is done in Fourier space, yielding essentially perfect interpolation of
sufficiently smooth flows when shifting a mode by any arbitrary fraction of a
grid cell. Two additional algorithms for reducing the space-dependence of the
numerical dissipation are presented in \App{s:xdepdiff}.

\subsection{Dissipation}
\label{s:dissipation}

The symmetric finite difference scheme of the Pencil Code has very little
intrinsic amplitude error (i.e.\ numerical diffusion) in the advection terms
\citep{Brandenburg2003}. Therefore we must add explicit dissipation terms to
the evolution equations in order to get rid of the kinetic and magnetic energy,
released from the gravitational potential by Reynolds and Maxwell stresses, on
the smallest scales of the grid.

\subsubsection{Viscosity}

The viscosity term $\vc{f}_\nu$ appears in full generality as
\begin{eqnarray}\label{eq:viscosity}
  \vc{f}_\nu &=& \nu_1
  \left[ \nabla^2\vc{u} + \frac{1}{3}\nab\nab\cdot\vc{u} +
  2(\vc{S}^{(1)}\cdot\nab\ln\rho)
  \right] \nonumber \\
  & &\hspace{-0.6cm} + \nu_3 \left[\nabla^6\vc{u}+(\vc{S}^{(3)}\cdot\nab\ln\rho)\right] \nonumber \\
  & &\hspace{-0.6cm}  + \nu_{\rm sh} \left[
  \nab\nab\cdot\vc{u}+(\nab\cdot\vc{u})(\nab\ln\rho)\right]
  + (\nab\nu_{\rm sh})\nab\cdot\vc{u} \, .
\end{eqnarray}
The viscosity includes regular Navier-Stokes viscosity with constant
coefficient $\nu_1$, hyperviscosity with constant coefficient $\nu_3$ and shock
viscosity with a variable coefficient $\nu_{\rm sh}$. In general we include
only a subset of the three types of viscosity, representing both direct
numerical simulations (DNS) and combined hyperviscosity/shock viscosity
solution of the equation of motion. In the following paragraphs we discuss the
three types of viscosity in more detail.

{\it Navier-Stokes viscosity} ($\nu_1$):
Since the density is not nearly constant in the stratified models, we must add
a density dependent term $2(\vc{S}^{(1)}\cdot\nab\ln\rho)$ to \Eq{eq:viscosity}
in order to conserve momentum. Here the traceless rate-of-strain tensor
$\vc{S}^{(1)}$ is
\begin{equation}
  S_{ij}^{(1)} = \frac{1}{2} \left( \frac{\dpa u_i}{\dpa x_j} + \frac{\dpa
  u_j}{\dpa x_i} - \frac{2}{3} \delta_{ij} \nab \cdot \vc{u} \right) \, .
\end{equation}

{\it Hyperviscosity} ($\nu_3$):
A simplified third order rate-of-strain tensor $\vc{S}^{(3)}$ is here defined as
\begin{equation}
  S_{ij}^{(3)} = \frac{\dpa^5 u_i}{\dpa x_j^5} \, .
\end{equation}
The high order Laplacian $\nabla^6$ in \Eq{eq:viscosity} is expanded as
$\nabla^6=\dpa^6/\dpa x^6+\dpa^6/\dpa y^6+\dpa^6/\dpa z^6$. This form of the
hyperviscosity conserves momentum \citep{JohansenKlahr2005}, but the energy
dissipation is not guaranteed to be positive by mathematical construction, nor
is the rate-of-strain tensor symmetric. We therefore compared all our results
with the evolution obtained by using a strict hyperviscosity operator
\citep[following][]{HaugenBrandenburg2004b} and found no qualitative
differences, so we opted to use the simplified hyperviscosity because it is
much simpler and quicker than the full version.

{\it Shock viscosity} ($\nu_{\rm sh}$):
The stratified runs produce shocks high up in the atmosphere of the disk. In
order to dissipate sufficient energy in the shocks, without rendering the whole
simulation domain strongly viscous, we have added an additional shock viscosity
term. The shock viscosity coefficient is obtained in a von Neumann-Richtmyer
fashion by taking the negative part of the divergence of $\vc{u}$, then taking
the maximum over three grid cells in each direction, and finally averaging over
three grid cells, to obtain
\begin{equation}\label{eq:nu_shock}
  \nu_{\rm sh}=c_{\rm sh} \langle {\rm max}[-\nab\cdot\vc{u}]_+ \rangle
  (\delta x)^{2} \, .
\end{equation}
We set the shock viscosity coefficient to $c_{\rm sh}=1.0$ in the stratified
runs to dissipate energy in shocks forming above two scale heights from the
mid-plane of the disk. The shock viscosity $\nu_{\rm sh}$ decreases rapidly
with increasing resolution, so convergence tests also probe the effect of
decreasing the shock viscosity term.
\begin{deluxetable*}{lccccccccc}
  \tabletypesize{\tiny}
  \tablecaption{Turbulence Properties}
  \tablewidth{0pt}
  \tablehead{
    \colhead{Run} &
    \colhead{$\langle\frac{1}{2}\rho u_x^2\rangle$} &
    \colhead{$\langle\frac{1}{2}\rho u_y^2\rangle$} &
    \colhead{$\langle\frac{1}{2}\rho u_z^2\rangle$} &
    \colhead{$\langle\frac{1}{2}B_x^2\rangle$} &
    \colhead{$\langle\frac{1}{2}B_y^2\rangle$} &
    \colhead{$\langle\frac{1}{2}B_z^2\rangle$} &
    \colhead{$\langle \rho u_x u_y\rangle$} &
    \colhead{$\langle -B_x B_y \rangle$} &
    \colhead{$\alpha$}
    }
  \startdata
    S           & $1.9\times10^{-3}$ & $1.7\times10^{-3}$ & $1.5\times10^{-3}$ & $1.2\times10^{-3}$ & $9.1\times10^{-3}$ & $5.3\times10^{-4}$ & $1.1\times10^{-3}$ & $4.3\times10^{-3}$  & $0.0036$ \\
    M           & $3.9\times10^{-3}$ & $5.1\times10^{-3}$ & $2.7\times10^{-3}$ & $4.2\times10^{-3}$ & $3.6\times10^{-2}$ & $2.1\times10^{-3}$ & $2.6\times10^{-3}$ & $1.2\times10^{-2}$ & $0.0097$ \\
    L           & $4.0\times10^{-3}$ & $4.5\times10^{-3}$ & $2.4\times10^{-3}$ & $3.7\times10^{-3}$ & $2.6\times10^{-2}$ & $2.0\times10^{-3}$ & $2.5\times10^{-3}$ & $1.1\times10^{-2}$ & $0.0089$ \\
    H           & $4.5\times10^{-3}$ & $4.2\times10^{-3}$ & $2.4\times10^{-3}$ & $3.5\times10^{-3}$ & $2.3\times10^{-2}$ & $1.8\times10^{-3}$ & $2.6\times10^{-3}$ & $1.0\times10^{-2}$ & $0.0085$ \\
    M\_r2       & $2.1\times10^{-3}$ & $2.8\times10^{-3}$ & $1.4\times10^{-3}$ & $2.3\times10^{-3}$ & $1.5\times10^{-2}$ & $1.2\times10^{-3}$ & $1.3\times10^{-3}$ & $6.5\times10^{-3}$ & $0.0052$ \\
    L\_nogz     & $3.2\times10^{-3}$ & $2.5\times10^{-3}$ & $1.4\times10^{-3}$ & $1.9\times10^{-3}$ & $1.3\times10^{-2}$ & $7.1\times10^{-4}$ & $1.8\times10^{-3}$ & $6.9\times10^{-3}$ & $0.0058$ \\
    \hline
    M\_Bz\_r0.5 & $5.8\times10^{-3}$ & $7.6\times10^{-3}$ & $4.0\times10^{-3}$ & $6.1\times10^{-3}$ & $6.6\times10^{-2}$ & $3.3\times10^{-3}$ & $3.8\times10^{-3}$ & $1.8\times10^{-2}$ & $0.0148$ \\
    M\_Bz       & $5.9\times10^{-3}$ & $1.0\times10^{-2}$ & $4.7\times10^{-3}$ & $1.1\times10^{-2}$ & $7.2\times10^{-2}$ & $6.8\times10^{-3}$ & $4.1\times10^{-3}$ & $2.5\times10^{-2}$ & $0.0196$ \\
    M\_Bz\_r2   & $5.7\times10^{-3}$ & $1.2\times10^{-2}$ & $5.0\times10^{-3}$ & $1.2\times10^{-2}$ & $6.7\times10^{-2}$ & $7.8\times10^{-3}$ & $4.0\times10^{-3}$ & $2.7\times10^{-2}$ & $0.0207$ \\
    \hline
    MS          & $2.9\times10^{-3}$ & $2.0\times10^{-3}$ & $1.2\times10^{-3}$ & $1.4\times10^{-3}$ & $1.0\times10^{-2}$ & $4.6\times10^{-4}$ & $1.5\times10^{-3}$ & $5.2\times10^{-3}$ & $0.0045$ \\
    MS\_r2      & $1.7\times10^{-3}$ & $1.2\times10^{-3}$ & $7.2\times10^{-4}$ & $9.1\times10^{-4}$ & $5.7\times10^{-3}$ & $3.4\times10^{-4}$ & $8.4\times10^{-4}$ & $3.2\times10^{-3}$ & $0.0027$ \\
    MS\_r2\_fix & $3.4\times10^{-3}$ & $2.1\times10^{-3}$ & $1.7\times10^{-3}$ & $1.4\times10^{-3}$ & $9.1\times10^{-3}$ & $6.2\times10^{-4}$ & $1.5\times10^{-3}$ & $5.3\times10^{-3}$ & $0.0046$ \\
    \hline
    SS\_r4\_nu1 & $6.5\times10^{-4}$ & $6.1\times10^{-4}$ & $4.2\times10^{-4}$ & $5.1\times10^{-4}$ & $6.3\times10^{-3}$ & $1.8\times10^{-4}$ & $4.0\times10^{-4}$ & $2.8\times10^{-3}$ & $0.0021$ \\
    SS\_r8\_nu1 & $1.0\times10^{-3}$ & $9.9\times10^{-4}$ & $7.1\times10^{-4}$ & $8.7\times10^{-4}$ & $8.9\times10^{-3}$ & $3.3\times10^{-4}$ & $6.6\times10^{-4}$ & $4.2\times10^{-3}$ & $0.0033$ \\
    MS\_r4\_nu1 & $1.9\times10^{-3}$ & $1.8\times10^{-3}$ & $8.7\times10^{-4}$ & $1.5\times10^{-3}$ & $1.4\times10^{-2}$ & $5.1\times10^{-4}$ & $1.2\times10^{-3}$ & $6.8\times10^{-3}$ & $0.0053$ \\
  \enddata
  \tablecomments{Col.\ (1): Name of run. Col.\ (2)-(4): Kinetic energy. Col.\
  (5)-(7): Magnetic energy. Col.\ (8): Reynolds stress. Col.\ (9): Maxwell
  stress. Col.\ (10): $\alpha$-value. Energies and stresses have been
  normalized with the mean thermal pressure in the box, $\langle P \rangle =
  c_{\rm s}^2 \langle \rho \rangle$.}
  \label{t:turbulence}
\end{deluxetable*}
\begin{figure*}
   \includegraphics[width=17.3cm]{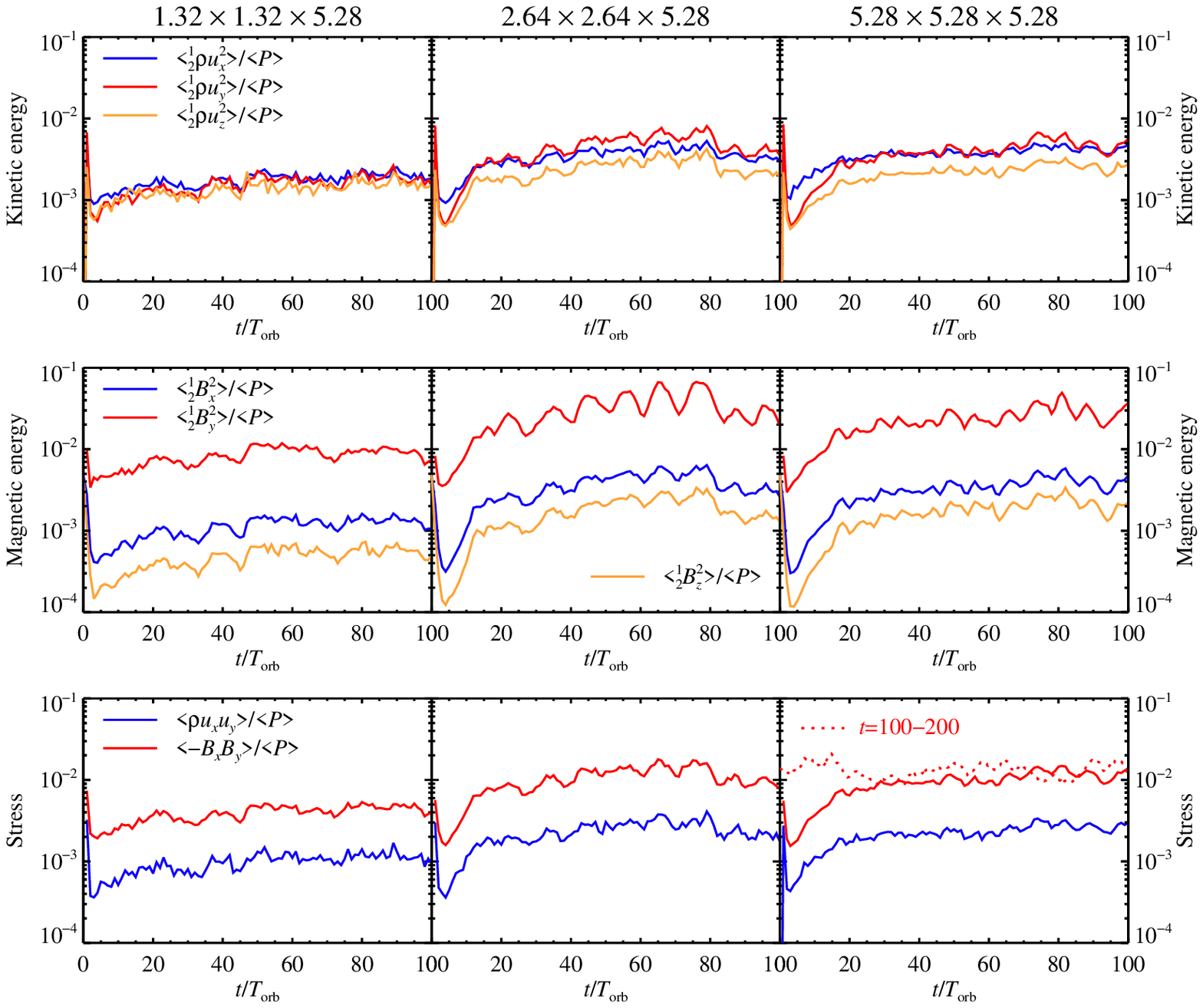}
   \caption{Kinetic energy (top row), magnetic energy (middle row) and
   turbulent stresses (bottom row) as a function of time measured in orbits,
   shown for three different box sizes along the columns (standard runs S, M
   and L). All quantities have been normalized by the mean thermal pressure in
   the box. Energies and stresses increase by more than a factor two when
   doubling the box dimensions from $L_x=L_y=1.32 H$ to $L_x=L_y=2.64 H$, but
   another doubling makes no significant difference for the statistical
   properties of the turbulence. The dotted line in the lower, right panel
   indicates the further evolution of the Maxwell stress from $t=100 T_{\rm
   orb}$ to $t=200T_{\rm orb}$, showing that the turbulent stresses in run L
   remain fairly constant after $t=30 T_{\rm orb}$.}
  \label{f:turbulence_properties}
\end{figure*}

\subsubsection{Resistivity}

For the resistivity term $\vc{f}_\eta$ we consider either normal resistivity or
hyperresistivity,
\begin{eqnarray}
  \vc{f}_\eta &=& \eta_1 \nabla^2\vc{\vc{A}} + \eta_3 \nabla^6 \vc{A} \, .
\end{eqnarray}
As for the hyperviscosity we use a simplified version for $\nabla^6$, but
checked that the strict high order Laplacian $\nabla^2\nabla^2\nabla^2$ gives
qualitatively and quantitatively similar results.

\subsubsection{Diffusion}

Mass diffusion consists of hyperdiffusion and shock diffusion
\begin{eqnarray}
  f_D &=& D_3 \nabla^6 \rho + D_{\rm sh} \nabla^2\rho+\nab D_{\rm
  sh}\cdot\nab\rho \, ,
  \label{eq:massdiff}
\end{eqnarray}
where $D_{\rm sh}=\nu_{\rm sh}$ is defined in \Eq{eq:nu_shock}. Mass diffusion
turned out to be necessary in order to damp out spurious small-scale modes that
are left behind due to the dispersion error in the finite differencing of the
advection term in the continuity equation. We also add a bit of extra mass
diffusion in the stratified simulations to stabilize the shocks forming above
two scale heights from the mid-plane of the disk (setting the shock diffusion
coefficient $D_{\rm sh}=1.0$ in Eq.\ [\ref{eq:massdiff}]).

\subsection{Boundary Conditions}

We use the usual shearing box boundary conditions in the radial and azimuthal
directions (shear-periodic/periodic). At the upper and lower boundaries we
apply periodic boundary conditions, conserving the average flux of all
components of the magnetic field. This turned out to be numerically less
challenging, and we checked that perfect conductor boundary conditions ($\dpa
B_x/dz=\dpa B_y/dz=B_z=0$) and vertical field boundary conditions ($B_x=
B_y=\dpa B_z/\dpa z=0$) for the magnetic field give results that are
qualitatively and quantitatively similar to what we see with periodic boundary
conditions.

\subsection{Simulation Parameters}

We run simulations for various box sizes, resolutions, dissipation parameters,
and imposed magnetic field strengths. The simulation parameters are given in
\Tab{t:runs}. The main simulations are the stratified S, M, L, and H runs.
These all share a vertical extent of $L_z=5.28 H$, but the radial/azimuthal
extent of the box is varied in steps of two, from $L_x=L_y=1.32$ in the
smallest box\footnote{The choice of $L=1.32$ as the basic box size unit is
based on the fact that $L_x=1.32$ approximately marks the transition from
subsonic to supersonic Keplerian shear flow.} (S) to $L_x=L_y=10.56$ in the
largest box (H). For the M run we perform a convergence test by doubling the
resolution (M\_r2), and we also present an unstratified version of run L
(L\_nogz), ignoring the $-\varOmega^2 z$ term in \Eq{eq:eqmot}. The two largest
box sizes (runs L, L\_nogz and H) are evolved using the interpolated Keplerian
shear scheme SAFI, described in \App{s:safi}. This is done in order to relax
the relatively tight time-step constraint of the Keplerian advection in these
large boxes and at the same time remove the numerical diffusivity associated
with the Keplerian advection.

The next group of simulations in \Tab{t:runs}, runs M\_Bz\_r0.5, M\_Bz,
M\_Bz\_r2, have a weak vertical magnetic field imposed on the gas, to keep the
turbulent activity relatively constant when increasing the resolution. This
makes it possible to better tell what effect resolution has on the zonal flows.
In the unstratified runs MS, MS\_r2 and MS\_r2\_fix, on the other hand,
turbulent activity is maintained when keeping the hyperdissipation parameters
constant as the resolution is increased.

The final set of simulations, runs SS\_r4\_nu1, SS\_r8\_nu1, MS\_r4\_nu1, apply
more natural second derivative viscosity and resistivity operators to dissipate
energy. A relatively high base resolution is needed here, as otherwise there
are not enough small scales for the second order operators to dissipate kinetic
and magnetic energy on. Strong zonal flows appear also with second derivative
dissipation terms, and we see relatively good convergence in the zonal flow
structure up to our highest resolution of around 200 grid points per scale
height (eight times higher than the S -- H runs).

\subsection{Units and Initial Conditions}

Throughout the paper we will state length in units of the scale height $H$,
velocity in units of the isothermal sound speed $c_{\rm s}$, time in units of
the inverse Keplerian frequency $\varOmega^{-1}$ or the orbital period $T_{\rm
orb}=2\pi\varOmega^{-1}$, density in units of the mid-plane density $\rho_0$
and magnetic field in units of $c_{\rm s} (\mu_0 \rho_0)^{1/2}$, related to the
square root of the thermal pressure. Energies and stresses are always
normalized by the mean thermal pressure in the box, $\langle P\rangle=c_{\rm
s}^2\langle \rho \rangle$.

In the stratified runs we set the initial condition for the density through an
isothermal hydrostatic equilibrium with scale height $H=c_{\rm s}/\varOmega$,
whereas the density is simply initialized to be constant in the non-stratified
runs. The toroidal component of the magnetic vector potential is set as a
moderate scale wave $A_y = A_0 \cos(k_x x) \cos(k_y y) \cos(k_z z)$ of
amplitude $A_0=0.04$ and wave number $k_x = k_y = k_z = 4.76 H^{-1}$, yielding
an initial plasma $\beta$ of around 50. We perturb the gas velocity components
with random fluctuations of amplitude $\delta u \sim 10^{-3} c_{\rm s}$.

\section{Turbulence Properties In Large Shearing Boxes}
\label{s:turbulence}

The measured properties of the turbulent velocity and magnetic fields are
written in \Tab{t:turbulence}. All quantities have been normalized by the
average pressure in the box, $\langle P \rangle = c_{\rm s}^2 \langle \rho
\rangle$, for better comparison between stratified and non-stratified runs. The
last column of \Tab{t:turbulence} indicates the $\alpha$-value of the flow
\citep{ShakuraSunyaev1973}, defined through the Reynolds and Maxwell stress as
\begin{equation}
  \alpha = \frac{2}{3} \frac{\langle (\rho u_x u_y-B_x B_y)\rangle}{\langle P
  \rangle} \, .
\end{equation}

The time evolution of kinetic energy, magnetic energy and turbulent stresses
for the standard runs S, M and L is plotted in \Fig{f:turbulence_properties}.
During the first orbit there is a sharp increase in both kinetic and magnetic
energies, followed by a rapid drop. After the drop the energies climb slowly
towards their saturation values, which are reached after around 20 orbits. In
the saturated state there is more than a doubling of the energies and stresses
when increasing the box size from $L_x=L_y=1.32$ (S) to $L_x=L_y=2.64$ (M), but
little difference between run M and the twice as large run L. The vertical
extent of the box is fixed at $L_z=5.28$ in all the simulations. The kinetic
energy is distributed relatively isotropically in the three components in the
small box. But a distinct split into a high radial/azimuthal contribution and a
low vertical contribution is seen when going to larger computational domains.
This split indicates a transition to scales dominated by the Coriolis force, as
geostrophic modes excited by the turbulence involve only in-plane gas motion.

The magnetic field is dominated by the azimuthal component for all domain
sizes, as expected since Keplerian shear stretches radial field and creates
azimuthal field of order $|\hat{B}_y| \sim (t_{\rm coh}/t_{\rm shear})
|\hat{B}_x|$, where $t_{\rm coh}$ is the coherence time of the radial field and
$t_{\rm shear}=(2/3)\varOmega^{-1}$ is the shear time-scale. The ratio ${\rm
rms}(B_y)/{\rm rms}(B_x) \approx 2.5$--$3.0$ for runs S--H gives a (scale
averaged) coherence time-scale of $t_{\rm coh}\approx2\varOmega^{-1}$. Based
on the long correlation time of large scale density and velocity structures
(\S\ref{s:zonalflows}), one might expect an even greater dominance of $B_y$,
especially in the larger boxes. However, the large scale zonal flow is built up
from small, uncorrelated kicks by the Lorentz force, and a relatively short
correlation time of the magnetic field fits well into this random walk picture.
\begin{figure}
   \includegraphics[width=8.7cm]{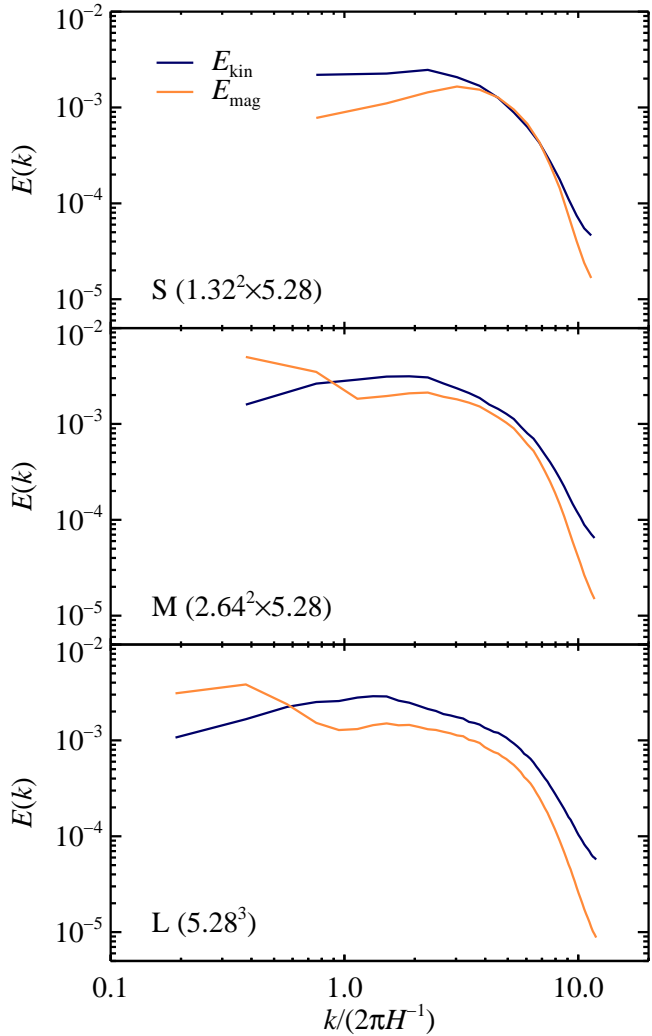}
   \caption{The kinetic (blue/dark) and magnetic (yellow/gray) energy spectrum,
   summed over shells of constant wavenumber $k=|\vc{k}|$. The kinetic energy
   dominates at the largest scales in the small box (top panel), whereas there
   is equipartition between kinetic and magnetic energy in a big range of
   smaller scales. In the two larger boxes (middle and bottom panels) the
   magnetic energy comes to dominate scales larger than approximately two scale
   heights. At the same time the equipartition at small scales is broken in
   favor of a dominance of kinetic energy.}
  \label{f:energy_spectrum}
\end{figure}

Maxwell and Reynolds stresses increase from $\langle \rho u_x u_y \rangle
\approx 0.001$ and $\langle -B_x B_y \rangle \approx0.004$ in the smallest box
to $\langle \rho u_x u_y \rangle \approx 0.0025$ and $\langle -B_x B_y \rangle
\approx0.01$ in the largest box, yielding an increase in $\alpha$-value from
$\alpha\approx0.004$ to $\alpha \approx0.009$. This increase is of interest
because global simulations of the MRI tend to yield higher $\alpha$-values than
traditional narrow shearing boxes
\citep{Hawley+etal1995,Brandenburg+etal1995,ArltRuediger2001,FromangNelson2006}.
Our simulations suggest a transition towards the global value with increasing
size of the shearing box. An additional explanation for the measured high
stresses in global simulations may be the relatively low resolution per scale
height possible in these simulations. The saturated state of non-stratified
zero net vertical flux magnetorotational turbulence has been shown to decline
with increasing resolution \citep{FromangPapaloizou2007}, due to the change in
numerical dissipation as the resolution increases. The turbulent transport
coefficients in run M\_r2, with twice as high resolution as run M, are
approximately 50\% lower than in run M (see \Tab{t:turbulence}). Such a linear
scaling of transport coefficients with resolution extends the effect found by
\cite{FromangPapaloizou2007} to stratified boxes.
\begin{figure*}
   \includegraphics[width=\linewidth]{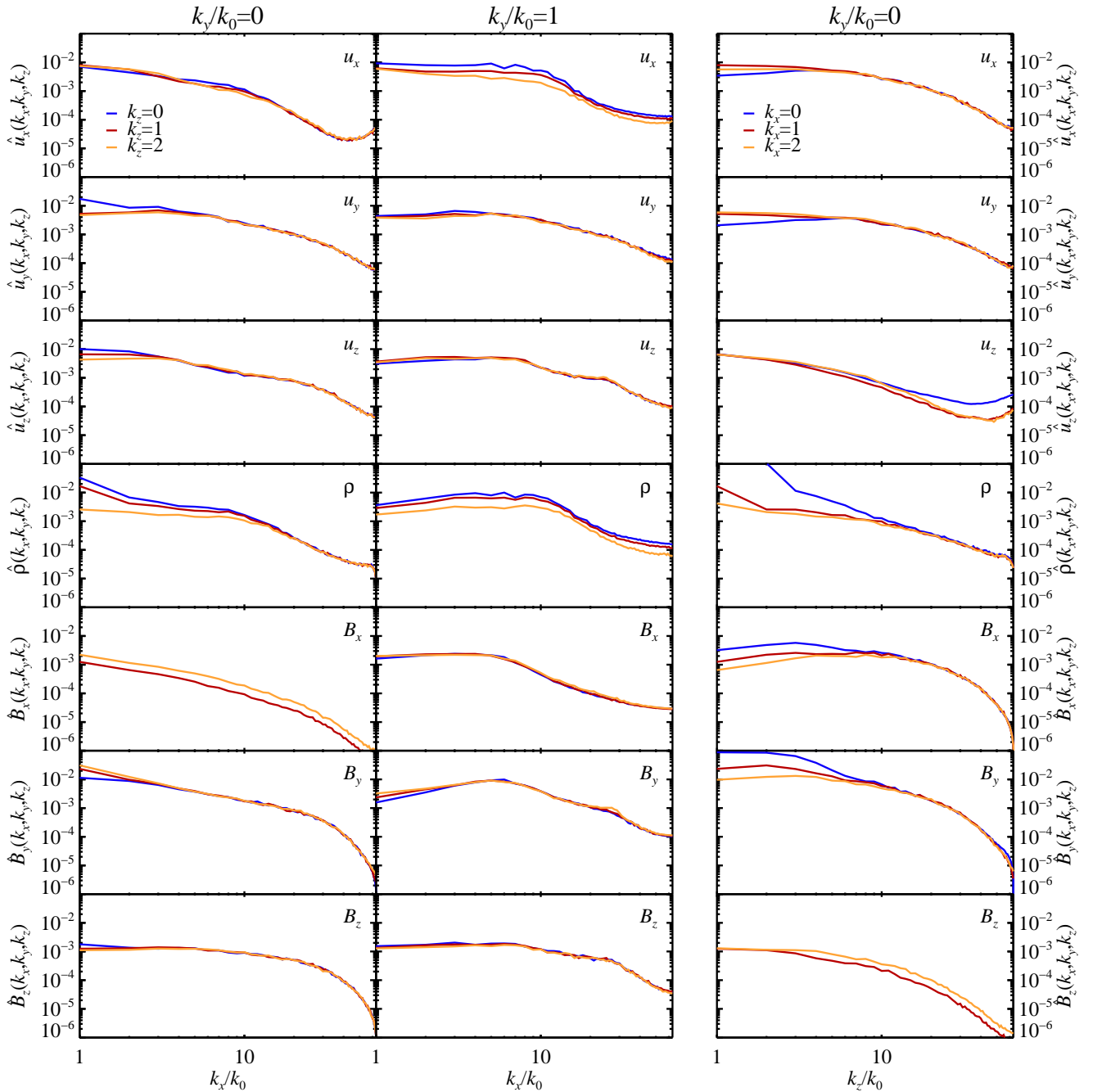}
   \caption{Fourier amplitudes of all dynamical variables for the stratified
   simulation with $L_x=L_y=L_z=5.28 H$, as a function of radial wavenumber
   $k_x$ (two first columns) and vertical wavenumber $k_z$ (last column).
   Colors indicate different wavenumbers in either the vertical direction (two
   first columns), or in the radial direction (last column), while the
   azimuthal wavenumber $k_y$ is indicated at the top of each column. There is
   a clear tendency for $u_x$, $u_y$, $\rho$, $B_x$ and $B_y$ to peak at the
   largest radial scale of the box. However the last column shows that most
   large scale magnetic energy resides in modes with vertical variation. The
   variation of $\rho$ with $z$ is mainly due to the stratification.}
  \label{f:fourier_amplitudes}
\end{figure*}

The kinetic and magnetic energy spectrum arising in boxes of different sizes is
shown in \Fig{f:energy_spectrum}, summed over wavenumber intervals
$|\vc{k}|\in[k_{\rm cen}-k_0/2,k_{\rm cen}+k_0/2$], with $k_{\rm cen}$ taking
integer values times the smallest wavenumber $k_0$. We have only included
isotropic scales in the summation. Thus we exclude (a) long vertical
wavelengths that exceed the radial or azimuthal extent of the smaller boxes in
runs S and M, and (b) the anisotropic high wavenumbers which only exist in the
``corners" of $k$-space. We averaged the energy spectra over 71 snapshots taken
between 30 and 100 orbits.

For the two large runs (M and L) the kinetic energy (blue/dark) line dominates
over magnetic energy (yellow/gray) at scales below approximately $1\ldots2$
scale heights. At larger scales (smaller $k$) the kinetic energy falls off
gradually, whereas the magnetic energy keeps rising until finally peaking near
the largest scales of the box. This transition from kinetically to magnetically
dominated flow can only be seen when considering a sufficiently large box size.
Interestingly the magnetic energy is comparable to kinetic energy near the
dissipative subrange in the smallest box, whereas the two energies move apart
when the box size is increased.
\begin{figure}
   \includegraphics[width=8.7cm]{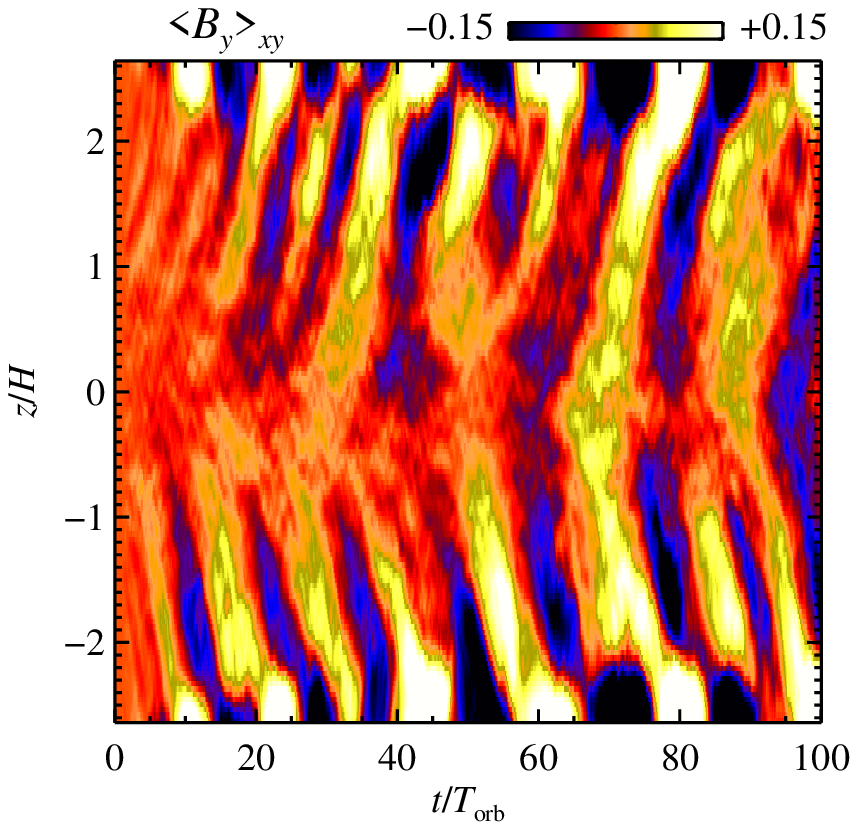}
   \caption{The azimuthal magnetic field $B_y$ of run L as a function of time
   and height over the mid-plane. Field structures continuously propagate away
   from the mid-plane, causing the azimuthal field at a given height over the
   mid-plane to flip signs with a period of approximately 13 orbits.}
  \label{f:butterfly}
\end{figure}

In \Fig{f:fourier_amplitudes} we show for run L the spectral power of all
dynamical components ($u_x$, $u_y$, $u_z$, $\rho$, $B_x$, $B_y$, $B_z$) as a
function of radial wavenumber ($x$-axis), the smallest two azimuthal
wavenumbers (columns) and the smallest three vertical wavenumbers (colored
lines) in the first two columns. The third column, on the other hand, shows the
spectral power as a function of vertical wavenumber, but for different, fixed
values of the radial wavenumber. Looking first at the magnetic field (three
bottom rows of \Fig{f:fourier_amplitudes}) one sees that the radial and
azimuthal magnetic energy peaks at the largest axisymmetric scales of the box
($k_y=0$), whereas the non-axisymmetric modes have far less power. However, the
large scale purely vertical modes shown in the third column contain
approximately a factor four times more magnetic energy than the radial modes.
The vertical component of the magnetic field has a rather flat dependence on
all directions and show little large scale excitation.

The azimuthal velocity field component and the gas density also exhibit very
high power at large axisymmetric scales; we return to those zonal flows in
\S\ref{s:zonalflows}.

\subsection{Magnetic Fields on Large Vertical Scales} \label{s:dynamo}

The large scale power of the $z$-dependent radial and azimuthal magnetic field
components, seen in the last column of \Fig{f:fourier_amplitudes}, is
illustrated in \Fig{f:butterfly}. In this figure we show the azimuthal magnetic
field, averaged over the $x$ and $y$ directions, as a function of time $t$ and
height over the mid-plane $z$. Magnetic field structures propagate away from
the mid-plane such that the field at a given height over the mid-plane flips
signs at regular intervals. Similar buoyancy waves are seen in the simulations
by \cite{Brandenburg+etal1995}, \cite{MillerStone2000} and
\cite{Hirose+etal2006}. The period of the sign flip is around 13 orbits, in
agreement with the turbulent diffusion time-scale $t_{\rm D}=H^2/\eta_{\rm t}$,
where $\eta_{\rm t}\sim0.01$ is the turbulent resistivity. This time is much
longer than the correlation time of less than one orbit deduced in the previous
section from the creation of azimuthal field from the radial field by the
Keplerian stretching term.

The vertically propagating radial and vertical field may be the effect of a
dynamo driven by rotation, stratification and shear
\citep{Parker1955,Moffatt1978}. However, by correlating the mean field with the
fluctuating electromotive force at both sides of the mid-plane,
\cite{Brandenburg+etal1995} concluded that the dynamo coefficient has the
opposite sign of what is expected from mean field dynamo theory.

The non-stratified simulation L\_nogz displays vertically dependent magnetic
field structures propagating from and to the mid-plane. Without stratification
there is no systematic helicity separation over the mid-plane, although patches
of positive and negative helicity appear with a correlation time of a few
orbits. The large-scale magnetic field activity may come around as a result of
an incoherent $\alpha$-$\omega$-dynamo \citep{VishniacBrandenburg1997},
thriving on the fluctuating helicity, or be connected to magnetorotational
instability in the azimuthal fields \citep{LesurOgilvie2008}.
\begin{figure*}
   \includegraphics[width=0.45\linewidth]{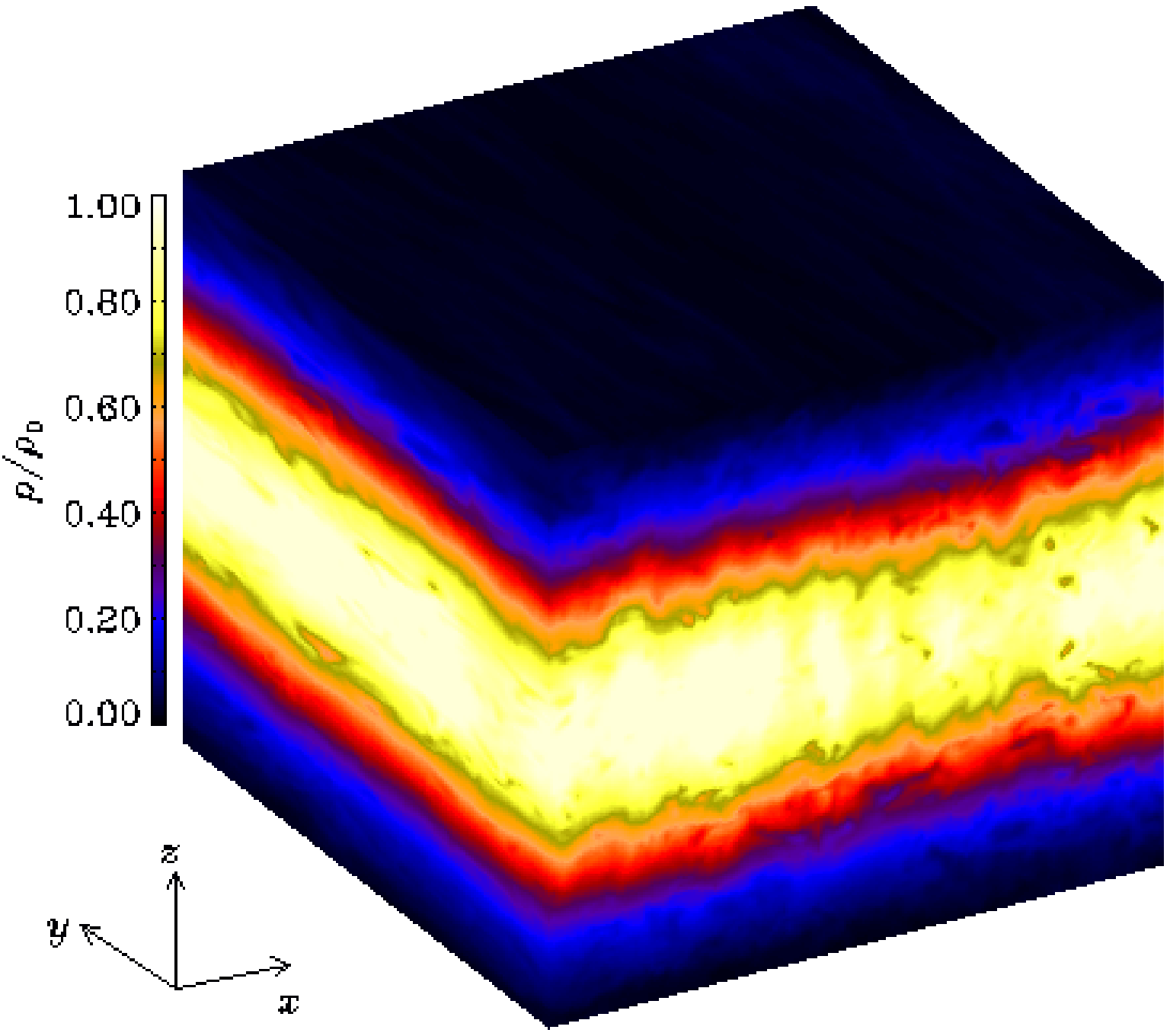}\qquad
   \includegraphics[width=0.45\linewidth]{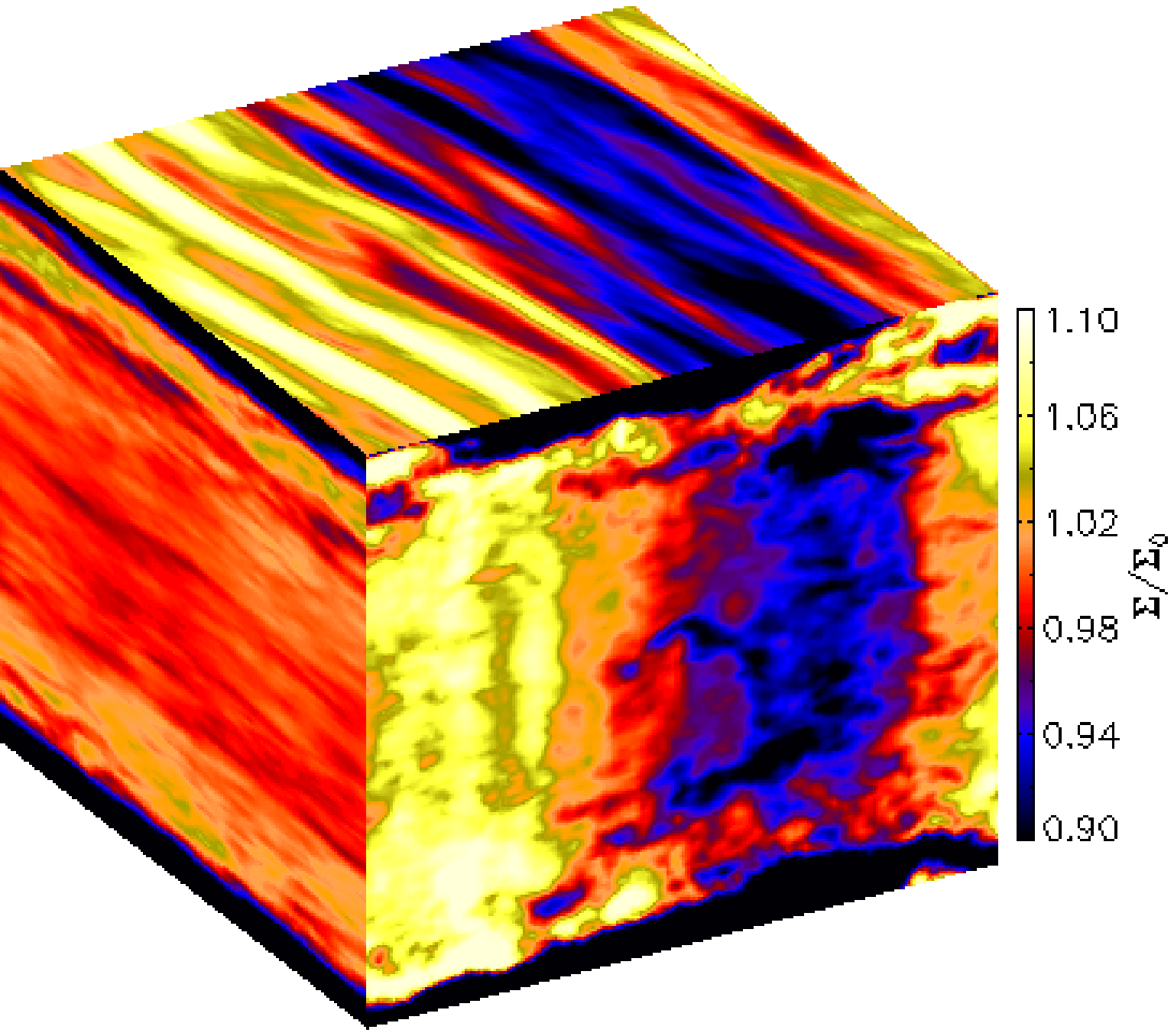}
   \caption{The gas density at the sides of the simulation box (left panel) and
   the directional column density (right panel, normalized by the column
   density of the original stratification) of run L at time $t=72 T_{\rm orb}$.
   The dominant column density structure is vertically constant and
   axisymmetric and has variation mainly at the largest radial scale of the
   box.}
  \label{f:boxes}
\end{figure*}
\begin{figure*}
   \includegraphics[width=\linewidth]{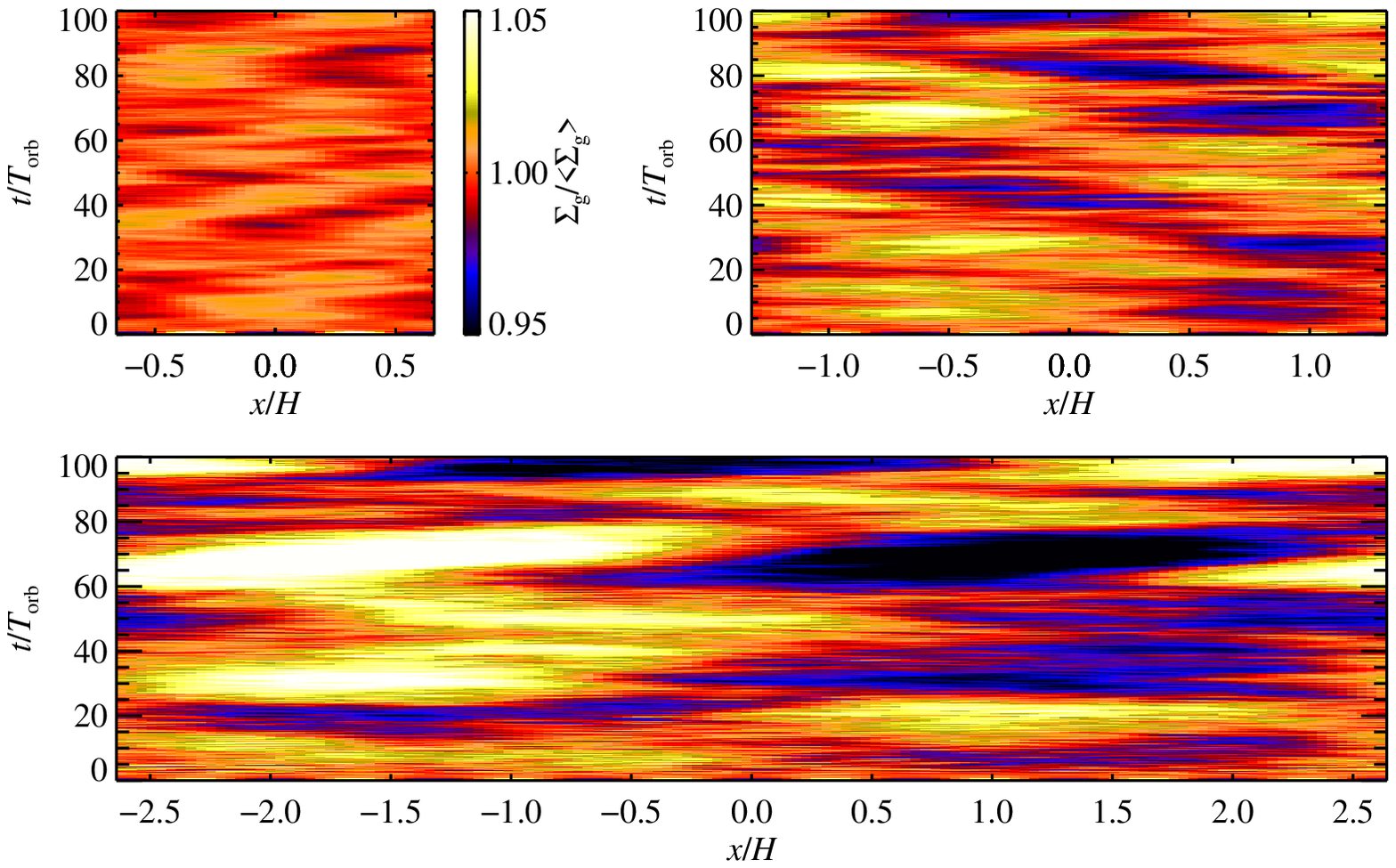}
   \caption{The gas density of runs S, M and L averaged over the $y$- and
   $z$-direction, as a function of radial distance from the center of the box,
   $x$, and the time, $t$, given in orbits. Density structures get higher
   amplitude and longer lived as the box size is increased from $L_x=L_y=1.32
   H$ to $L_x=L_y=5.28 H$. The height of the box stays fixed at $L_z=5.28 H$.}
  \label{f:rhomx_t}
\end{figure*}

\section{Zonal Flows}
\label{s:zonalflows}

In \Fig{f:boxes} we show the gas density at the sides of the simulation box at
$t=72\,T_{\rm orb}$ for run L (box size of $5.28 H\times5.28 H\times5.28 H$).
The turbulent structure of the gas density is clearly seen superimposed on the
main stratification in the left panel. The right panel shows the column density
along each direction $\varSigma_i=\int \rho(x,y,z) \de x_i$, divided by the
directional column density in the initial stratification. There is a dominant
column density structure at the largest radial scale of the box, with very
little dependence in the azimuthal and vertical directions\footnote{The marked
drop in column density near the upper and lower boundary planes is likely due
to the periodic boundary conditions that precludes hydrostatic equilibrium near
the boundaries by forcing the density derivative to be zero.}.

The space-time plots of the gas density of runs S, M and L shown in
\Fig{f:rhomx_t} confirm the dominance of the $k_x/k_0=1$ mode. Here we have
averaged over the vertical and azimuthal directions and show the gas density as
a function of radial coordinate $x$ and time $t$ measured in orbits. Both the
density amplitude and the correlation time of the large scale structures
increases significantly when making the box size larger. Run H is not shown in
\Fig{f:rhomx_t} -- the long correlation time of the density structures in this
huge box would require several hundred orbits integration to get good
statistics.

\begin{figure}
   \includegraphics[width=\linewidth]{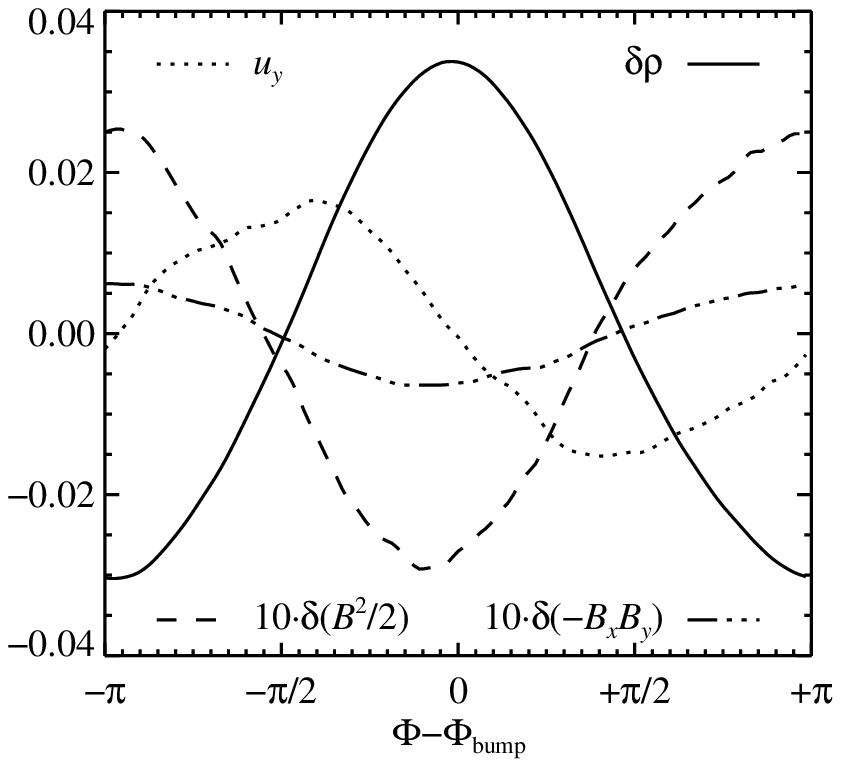}
   \caption{Phase-shifted and time-averaged perturbations to the gas density,
   azimuthal velocity, Maxwell stress (multiplied by 10 for visibility) and
   magnetic pressure (also multiplied by 10) for run L, as a function of the
   radial phase $\Phi$. The $-\pi/2$ phase shift of $u_y$ is a consequence of
   geostrophic balance on large scales. There is a clear anti-correlation of
   gas pressure (proportional to density) with Maxwell stress and magnetic
   pressure. Perturbations are azimuthally and vertically averaged for each
   time-step. Prior to time-averaging, the quantities are shifted in radius so
   that the lowest order radial perturbation (with $k_x=k_0=2\pi/L_x$) to the
   gas density is centered in the plot.}
  \label{f:phasesum}
\end{figure}

Axisymmetric density modes have their power peak at the largest radial scale of
the box. The azimuthal velocity has a similar peak at the largest axisymmetric
modes (see \Fig{f:fourier_amplitudes}). \Fig{f:phasesum} compares the
large-scale behavior of the density fluctuations to other physical quantities.
Quantities are radially shifted so that the $k_x = k_0$ density bump peaks at
the plot center, and then are time averaged. This filtering causes the plotted
quantities (especially the gas density) to appear sinusoidal on the box scale,
since higher order modes do not have a consistent phase relationship with the
largest mode. 
 
\Fig{f:phasesum} shows an almost exact $-\pi/2$ phase difference between the
radial peaks of the density and azimuthal velocity. This follows from near
geostrophic balance between Coriolis forces and pressure,
\begin{equation} 
  2 \rho_0 \varOmega u_y \approx \dpa P/\dpa x\, ,
\end{equation} 
which gives super-Keplerian (sub-Keplerian) motions radially interior
(exterior) to the density peak. In detail though the radial force balance
includes Lorentz forces. The magnetic pressure and Maxwell stress perturbations
in \Fig{f:phasesum} are almost perfectly anti-correlated with the thermal
pressure (the trend for thermal and magnetic pressures to misalign comes
naturally out of the simplified zonal flow model presented in
\S\ref{s:zonalflowmodel}). This supports recent findings by T.\ Sano (personal
communication, 2008).

Azimuthally and vertically varying modes have been averaged out of the density
structures that appear in \Fig{f:rhomx_t}. But \Fig{f:fourier_amplitudes} shows
that there is a clear trend for the density power to fall quickly with
azimuthal wavenumber, likely because the Keplerian shear drives
non-axisymmetric spectral power towards larger and larger $k_x$. The decreased
spectral power of density at vertically varying modes can be seen as an effect
of the Taylor-Proudman theorem. There is no geostrophic balance available for
modes with vertical variation other than the main stratification, so pressure
and turbulent diffusion will even out any vertical variation faster than the
magnetorotational instability injects energy to those modes.

In \Tab{t:zonalflow} we show the measured values for the Fourier amplitudes of
azimuthal velocity, density and Maxwell stress at the largest radial scale of
the box, the power law index for the two largest radial scales, and the
correlation time of the density structures. The power law index is found from
\begin{equation}
  \zeta = \frac{\ln[|\hat{f}|(2 k_0,0,0)]-\ln[|\hat{f}|(k_0,0,0)]}{\ln 2} \, .
\end{equation}
We have normalized the density amplitude by the mean density in the box to get
effectively the column density perturbation relative to the mean column
density, $\hat{\rho}(k_0,0,0)/\langle\rho\rangle =
\hat{\varSigma}(k_0,0,0)/\langle\varSigma\rangle$. The correlation time has
been calculated by taking the value of $\rho$, averaged over $y$ and $z$, at a
given time $t$ and measuring for each spatial point the time it takes for the
density at that point to change by a value corresponding to the standard
deviation of the gas density. The result has been averaged over many closely
spaced times (for which the turbulence is saturated and the correlation time
does not extend to the final time of the simulation) and multiplied by two to
represent the full temporal extent of the correlated structures. The resulting
correlation times are in good agreement with the life-time of the structures
seen \Fig{f:rhomx_t}, and the measurements confirm the very long correlation
times seen in simulation L. For the even larger box, simulation H, the
correlation time is approximately 50 orbits, comparable to the total simulation
time.

Interestingly the amplitude of the large scale azimuthal velocity is relatively
independent of box size, while the density amplitude grows approximately
proportional to the radial extent of the box\footnote{Parceval's theorem would
imply that the power at a given Fourier mode is inversely proportional to the
number of grid points, but for the case of zonal flows, the power peaks so
strongly at the largest radial mode that this reduction does not apply.}.
There is an almost perfect geostrophic connection between the amplitude of the
zonal flow and the normalized density amplitude, following the expected
$|\hat{u}_y|\sim(1/2) k_0 |\hat{\rho}|$. An exception to this is the mean
vertical field runs (M\_Bz\_*) where the azimuthal velocity is higher than
needed for geostrophic balance. We give a possible explanation for this
discrepancy in \S\ref{s:complete}, based on the relatively short correlation
time of the turbulence in the imposed field runs.
 
The power law index of azimuthal velocity and density also follows the
geostrophic expectation of $\zeta_\rho=\zeta_{u_y}-1$, although both
$\zeta_{u_y}$ and $\zeta_\rho$ decrease (towards zero) with increasing box
size. This decrease could indicate some motion towards convergence of the zonal
flow and pressure bumps. Global simulations will be needed to eventually find
such convergence, since curvature effects could play an important role for
simulation domains of more than ten scale heights.

The presence of large scale radial variation in Maxwell stresses is clear from
\Tab{t:zonalflow}. The amplitude is typically around 10--15\% of the mean
Maxwell stress, with a sharp decrease towards shorter wavelengths
($\zeta_{B_xB_y}\approx-1$). The imposed field simulations (M\_Bz\_*) even
indicate an increasingly top heavy Maxwell stress with increasing resolution.
However, the turbulence activity in that set of runs also increases by 50\% as
the resolution is increased (see \Tab{t:turbulence}), while the amplitude of
the large scale Maxwell stress increases somewhat slower.

\subsection{Large-Scale Balance}

\begin{deluxetable*}{lcccccccr}
  \tabletypesize{\small}
  \tablecaption{Zonal Flow Properties}
  \tablewidth{0pt}
  \tablehead{
    \colhead{Run} &
    \colhead{$\langle -B_x B_y \rangle$} &
    \colhead{$|\hat{u}_y|(k_0,0,0)$} &
    \colhead{$|\hat{\rho}|(k_0,0,0)$} &
    \colhead{$|\widehat{B_x B_y}|(k_0,0,0)$} &
    \colhead{$\zeta_{u_y}$} &
    \colhead{$\zeta_{\rho}$} &
    \colhead{$\zeta_{B_x B_y}$} &
    \colhead{$t_{\rm corr}$}
    }
  \startdata
    S          & $4.3\times10^{-3}$ & $1.9\times10^{-2}$ & $8.1\times10^{-3}$ &
                 $4.5\times10^{-4}$ & $-1.6$ & $-2.4$ & $-0.9$ &  $8.3$ \\
    M          & $1.2\times10^{-2}$ & $2.3\times10^{-2}$ & $1.9\times10^{-2}$ &
                 $1.2\times10^{-3}$ & $-1.0$ & $-1.8$ & $-0.8$ &  $8.5$ \\
    L          & $1.1\times10^{-2}$ & $1.7\times10^{-2}$ & $3.3\times10^{-2}$ &
                 $1.3\times10^{-3}$ & $-1.0$ & $-2.3$ & $-1.4$ & $19.6$ \\
    H          & $1.0\times10^{-2}$ & $1.5\times10^{-2}$ & $5.2\times10^{-2}$ &
                 $1.1\times10^{-3}$ & $-0.2$ & $-1.1$ & $-0.9$ & $50.5$ \\
    M\_r2      & $6.5\times10^{-3}$ & $1.6\times10^{-2}$ & $1.3\times10^{-2}$ &
                 $5.5\times10^{-4}$ & $-0.9$ & $-2.2$ & $-1.0$ & $12.6$ \\
    L\_nogz    & $6.9\times10^{-3}$ & $1.2\times10^{-2}$ & $2.1\times10^{-2}$ &
                 $5.6\times10^{-4}$ & $-0.8$ & $-1.7$ & $-1.0$ & $18.8$ \\
    \hline
    M\_Bz\_r0.5& $1.8\times10^{-2}$ & $4.5\times10^{-2}$ & $2.4\times10^{-2}$ &
                 $2.1\times10^{-3}$ & $-1.1$ & $-1.5$ & $-0.6$ &  $4.6$ \\
    M\_Bz      & $2.5\times10^{-2}$ & $3.9\times10^{-2}$ & $2.4\times10^{-2}$ &
                 $2.4\times10^{-3}$ & $-0.6$ & $-1.9$ & $-0.9$ &  $5.4$ \\
    M\_Bz\_r2  & $2.7\times10^{-2}$ & $3.7\times10^{-2}$ & $1.8\times10^{-2}$ &
                 $2.5\times10^{-3}$ & $-0.5$ & $-1.5$ & $-1.2$ &  $3.2$ \\
    \hline
    MS         & $5.2\times10^{-3}$ & $2.7\times10^{-2}$ & $2.3\times10^{-2}$ &
                 $9.6\times10^{-4}$ & $-1.1$ & $-1.5$ & $-0.9$ &  $9.9$ \\
    MS\_r2     & $3.2\times10^{-3}$ & $1.5\times10^{-2}$ & $1.3\times10^{-2}$ &
                 $4.0\times10^{-4}$ & $-0.8$ & $-1.3$ & $-0.8$ & $11.3$ \\
    MS\_r2\_fix& $5.3\times10^{-3}$ & $2.9\times10^{-2}$ & $2.5\times10^{-2}$ &
                 $7.7\times10^{-4}$ & $-0.9$ & $-1.4$ & $-0.5$ & $15.7$ \\
    \hline
    SS\_r4\_nu1& $2.8\times10^{-3}$ & $2.0\times10^{-2}$ & $8.5\times10^{-3}$ &
                 $3.9\times10^{-4}$ & $-1.3$ & $-2.1$ & $-0.8$ &  $6.5$ \\
    SS\_r8\_nu1& $4.2\times10^{-3}$ & $2.2\times10^{-2}$ & $9.4\times10^{-3}$ &
                 $5.4\times10^{-4}$ & $-1.5$ & $-2.3$ & $-0.9$ &  $7.6$ \\
    MS\_r4\_nu1& $6.8\times10^{-3}$ & $3.3\times10^{-2}$ & $2.8\times10^{-2}$ &
                 $9.3\times10^{-4}$ & $-1.7$ & $-2.4$ & $-1.0$ &  $9.4$ \\
  \enddata
  \tablecomments{Col.\ (1): Name of run. Col.\ (2): Maxwell stress, normalized
  by mean pressure. Col.\ (3)-(5): Fourier amplitude of azimuthal velocity,
  density normalized by mean density in the box, and pressure-normalized
  Maxwell stress, respectively, at largest radial mode. Col.\ (6)-(8): Power
  law index of azimuthal velocity, density, and Maxwell stress, respectively,
  calculated from two largest radial modes. Col.\ (9): Correlation time, in
  orbits $T=2\pi\varOmega^{-1}$, of largest radial density mode.}
  \label{t:zonalflow}
\end{deluxetable*}
\begin{figure*}
   \includegraphics[width=\linewidth]{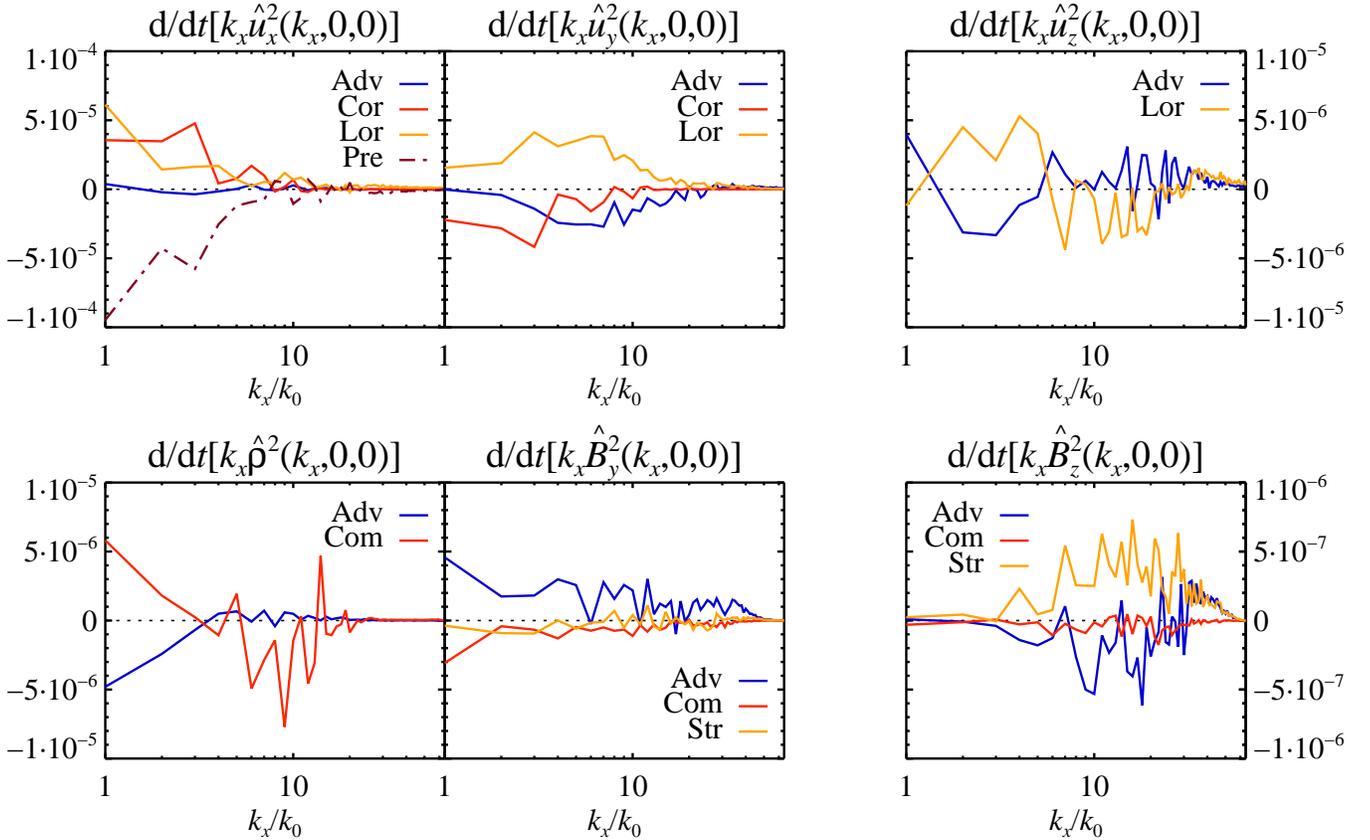}
   \caption{Power budget for all axisymmetric and vertically constant modes in
   run L. The curves show the power contribution from individual terms in the
   dynamical equations [``Adv''=advection, ``Cor''=Coriolis force,
   ``Lor''=Lorentz force, ``Pre''=pressure gradient, ``Com''=compression,
   ``Str''=stretching]. The data has been averaged over equally spaced
   snapshots from $t=30 T_{\rm orb}$ to t=$100 T_{\rm orb}$ and multiplied by
   $k_x$ for better visibility of the small scale power. The energy in the
   azimuthal velocity field (i.e.\ the zonal flow) is pumped by the Lorentz
   force (magnetic tension). The resulting convergence in the radial velocity
   leads to an increase in density amplitude and a decrease in azimuthal
   magnetic field. The compressive increase in gas density is in turn balanced
   by turbulent diffusion. The non-linear advection term in the induction
   equation for the azimuthal field, on the other hand, adds magnetic energy at
   the largest scale.}
  \label{f:energy_balance_all}
\end{figure*}
To get an understanding of the launching mechanism for zonal flows, we plot in
\Fig{f:energy_balance_all} the energy balance of all vertically constant and
axisymmetric modes. We have taken each term $Q_i$ in the dynamical equation of
variable $f$, $\dot{f}=\Sigma_i Q_i$, and calculated its contribution to the
power of all the purely radial modes as
\begin{equation}
  \frac{\dpa |\hat{f}|^2}{\dpa t} = \sum_i 2 {\rm Re}(\hat{f}^*\hat{Q}_i) \, .
  \label{eq:power}
\end{equation}
Here hats denotes Fourier transforms and stars complex conjugation. We add the
power contribution from both the negative and the positive wavenumber in
\Fig{f:energy_balance_all}. The second panel of \Fig{f:energy_balance_all}
indicates that the zonal flow [$\hat{u}_y(k_x)$] is excited primarily by the
Lorentz force (indicated by the positive value of the ``Lor'' curve). A large
scale radial variation in the Maxwell stress (see \Tab{t:zonalflow} and
\S\ref{s:cascade}) exerts a significant azimuthal tension on the gas. As the
zonal flow is launched by the magnetic tension, the Coriolis force creates a
slight radial component of the velocity field. The ensuing convergence of the
gas flow creates the radial pressure bumps seen in \Fig{f:rhomx_t} in perfect
geostrophic balance between Coriolis force and pressure gradient force.

\begin{figure}
   \includegraphics[width=8.7cm]{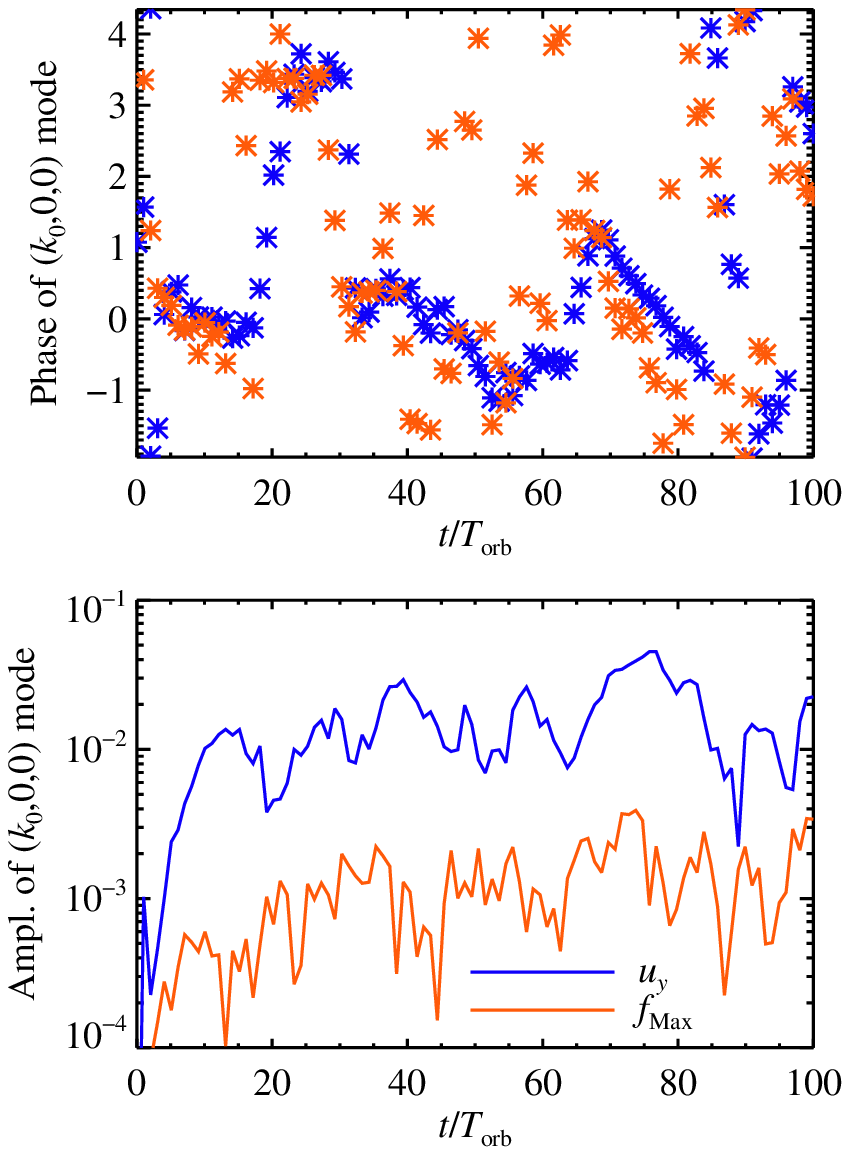}
   \caption{Phase and amplitude of the large-scale mode
   $(k_x,k_y,k_z)/k_0=(1,0,0)$, for the azimuthal component of the velocity
   field and the magnetic tension associated with the Maxwell stress.  There is
   some correlation between the phases (top panel), corroborating that zonal
   flows are indeed launched by the $x$-dependent Maxwell stress, although the
   Maxwell stress has variation on a much shorter time-scale than the zonal
   flows. The amplitude of the magnetic tension is approximately an order of
   magnitude smaller than the azimuthal velocity (bottom panel).}
  \label{f:phase_ampl_flory_uy_t}
\end{figure}
\Fig{f:phase_ampl_flory_uy_t} shows that there is some correlation between the
phase of the zonal flow and the phase of magnetic tension associated with the
Maxwell stress, although the latter varies on a much shorter time-scale than
the zonal flow. The magnetic tension is only an order of magnitude lower in
amplitude than the saturated zonal flow, so a correlation time of the magnetic
field of a few orbits is enough to explain the build up of large scale flow.

\section{Simplified Model of Zonal Flows}
\label{s:zonalflowmodel}

The zonal flows in our simulations have a very long correlation time, on the
order of many tens of orbits. We can understand both the amplitude and the
life-time of the zonal flows by adopting a simplified dynamical model which is
forced by the azimuthal magnetic tension. Consider the following set of
equations for the evolution of the radial velocity $\hat{u}_x$, azimuthal
velocity $\hat{u}_y$ and density $\hat{\rho}$, all for the large scale $\vc{k}
= (k_0,0,0)$ mode,
\begin{eqnarray}
  0 &=&
    2 \varOmega \hat{u}_y - \frac{c_{\rm s}^2}{\rho_0}
    \ii k_0 \hat{\rho} \label{eq:duxdteq} \, , \\
  \frac{\de \hat{u}_y}{\de t} &=&
    -\frac{1}{2} \varOmega \hat{u}_x + \hat{T} \, , \label{eq:duydteq} \\
  \frac{\de \hat{\rho}}{\de t} &=& -\rho_0 \ii k_0 \hat{u}_x -
  \frac{1}{\tau_{\rm mix}} \hat{\rho} \, . \label{eq:dlnrhodteq}
\end{eqnarray}
Equation (\ref{eq:duxdteq}) expresses radial geostrophic
balance\footnote{Dropping $\de \hat{u}_x/\de t$ filters high frequency density
waves which are not relevant for the longer-lived structures considered here.}.
We ignore for simplicity the role of Lorentz forces in this balance, evident
from \Fig{f:energy_balance_all} to be an order unity correction. The azimuthal
force balance in \Eq{eq:duydteq} describes forcing by the azimuthal tension
$\hat{T}$, which is partly deflected into radial motion, but also directly
accelerates the zonal flow in $\hat{u}_y$. Mass continuity in
\Eq{eq:dlnrhodteq} includes a damping term with the timescale $\tau_{\rm mix}$.
This represents non-linear advection terms, seen to be significant in
\Fig{f:energy_balance_all} (bottom left panel). The same figure (top left and
top center panels) shows that non-linear advection makes a negligible
contribution to the momentum equations.

We can combine the above equations into a single evolution equation for the
density,
\begin{equation}\label{eq:drhohatdt}
  \frac{\de \hat{\rho}}{\de t} = \frac{1}{1+ k_0^2 H^2 } \left( \hat{F}
    - \frac{\hat{\rho}(t)}{\tau_{\rm mix}} \right) \, ,
\end{equation}
where $ \hat{F} \equiv - 2 \ii k_0 \rho_0\hat{T}/\varOmega$ is the forcing
term. The prefactor $c_k\equiv(1 + k_0^2 H^2)^{-1}$ is a pressure correction
for small-scale modes that both decreases the amplitude of the forcing and
increases the effective damping time. The density bump grows in exact antiphase
with the stress bump $\hat{S}$, equivalent to the $\varSigma S={\rm const}$
relation of steady global accretion \citep{Pringle1981}. This follows from
writing $\hat{F} = -2 k_0^2 \varOmega^{-1} \hat{S}$, where the complex stress
amplitude is defined as $\hat{S} = -\mu_0^{-1} \widehat{B_x B_y}$.

The straightforward equilibrium solution to \Eq{eq:drhohatdt} is
$\hat{\rho}_{\rm eq}=\tau_{\rm mix} \hat{F}$. This is however not the correct
solution to the problem since \Fig{f:phase_ampl_flory_uy_t} clearly
demonstrates that the magnetic tension forcing $\hat{T}$ does not maintain a
consistent phase relationship with $\hat{u}_y$ (and thus with $\hat{\rho}$)
during the life-time of a zonal flow.

Instead \Eq{eq:drhohatdt} should be modeled as a stochastic differential
equation that gives the growth of the density and azimuthal velocity as a
damped random walk. We now give a description of the approximate
solution\footnote{More formal solution methods exist as a manifestation of the
fluctuation dissipation theorem. \citet{YoudinLithwick2007} consider a
mathematically similar problem: the stirring of particle velocities by
aerodynamic coupling to turbulent gas. They show that the approximate arguments
given here are consistent with more formal solutions to the stochastic Langevin
equation via Fokker-Planck analysis.}. We assume that the forcing term $c_k
\hat{F}$ has a correlation time $\tau_{\rm for}$. The density amplitude changes
during each coherent step by $|\hat{\rho}_{\rm step}| \sim c_{k} |\hat{F}|
\tau_{\rm for}$, and grows as a random walk, $ \hat\rho \sim |\hat{\rho}_{\rm
step}| \sqrt{N_{\rm step}} \sim |\hat{\rho}_{\rm step}| \sqrt{t/\tau_{\rm
for}}$. The damping term sets the lifetime of coherent structures as
\begin{equation}
  \tau_{\rm corr} \sim \tau_{\rm mix}/c_k \sim (1 + k_0^2H^2)/(k_0^2 D_{\rm
  t})\, ,
  \label{eq:taucorr}
\end{equation} 
where the final step describes the mixing process as diffusive with a strength
$D_{\rm t}$. The random walk saturates after $N_{\rm step} \sim \tau_{\rm
corr}/ \tau_{\rm for}$ steps, resulting in the equilibrium value
\begin{equation}\label{eq:rhoeq}
  \frac{\hat{\rho}_{\rm eq}}{\rho_0} = 2 \sqrt{c_k \tau_{\rm for}\tau_{\rm mix}}
  H k_0 \frac{\hat{T}}{c_{\rm s}} \, .
\end{equation}

We appeal to the simulations for appropriate values of $\hat{T}$, $\tau_{\rm
mix}$ and $\tau_{\rm for}$ to check the consistency of the model. The forcing
by magnetic tension can be estimated for run L from
\Fig{f:phase_ampl_flory_uy_t} to be approximately $\hat{T}\sim0.001\varOmega
c_{\rm s}$. From the same figure we read the correlation time of the forcing to
be approximately $\tau_{\rm for}\sim10\varOmega^{-1}$ (around two orbits). We
assess the turbulent mixing time-scale as a diffusive process with $\tau_{\rm
mix}=1/(k_0^2 D_{\rm t})\sim70\varOmega^{-1}$. Here we use $D_{\rm t}\sim\alpha
H^2 \varOmega \sim 0.01 H^2 \varOmega$ for the turbulent diffusion coefficient.

With the above quantities, \Eq{eq:rhoeq} gives an equilibrium density amplitude
$\hat{\rho}\sim0.041$ for run L, fairly close to the measured value
$\hat{\rho}\sim0.033$. This agreement is remarkable for such a simple model and
actually a bit fortuitous, since \Fig{f:energy_balance_all} (top left panel)
shows that radial Lorentz forces (neglected in this toy model) balance a
significant portion of the pressure gradient, requiring a yet higher density
perturbation.

\subsection{Life-times of Pressure Bumps and Zonal Flows}
\label{s:lifetime}

Zonal flow lifetimes are a useful constraint on our understanding of large
scale dynamics and mixing in MRI turbulence. The correlation times in
\Tab{t:zonalflow} can be compared to the simple predictions of the random walk
model in \Eq{eq:taucorr}\footnote{Where $t_{\rm corr} \equiv \tau_{\rm
corr}/(2\pi)$ since the former is in units of $T_{\rm orb}$.}. In large
shearing boxes, with $k_0 H\ll1$, the lifetime increases with the diffusive
mixing timescale, i.e.\ $t_{\rm corr} \propto L_x^2$. For smaller boxes, with
$k_0 H \gg 1$, the predicted $t_{\rm corr}$ is independent of box size for a
fixed diffusion coefficient.

Indeed, $t_{\rm corr} \approx 8~T_{\rm orb}$ is nearly constant in the two
smallest boxes, S and M, in rough agreement with the simple model. For runs L
and H our estimate gives $t_{\rm corr}\propto L_x^{1.23}$ in good agreement
with measured value of  $t_{\rm corr}\propto L_x^{1.37}$ for these runs.

\subsection{Consistency Checks on Forcing Model}\label{sec:consistency}

We now check the consistency of our model further, first by computing the
expected forcing timescale (the least well-measured quantity) in terms of
better measured quantities. Inverting \Eq{eq:rhoeq} gives
\begin{equation} 
  \tau_{\rm for} \approx \left({\hat{\rho} \over 2 k_0 c_k \hat{T}}\right)^2{1
  \over \tau_{\rm corr}}\, ,
\end{equation} 
where we use $H = c_{\rm s} = \rho_0 = 1$ and $\tau_{\rm corr} = 2\pi \varOmega
t_{\rm corr} = \tau_{\rm mix}/c_k$. Since these are all known quantities (using
$\hat{T} \propto k_0 \widehat{B_xB_y}$ normalized to $\hat{T} = 10^{-3}$ in run
L), we compute $t_{\rm for} = \tau_{\rm for}/(2\pi) \approx
\{0.64,0.62,1.5,9.8\}$ orbits for runs \{S,M,L,H\}. That $t_{\rm for} \ll
t_{\rm corr}$ supports the basic assumption of a random walk model.

As another consistency check, we can measure the turbulent mixing coefficient,
$D_{\rm t} \equiv (\tau_{\rm mix} k_0^2)^{-1}$ using the result from the random
walk model that $2 \pi t_{\rm corr} \approx \tau_{\rm mix}/c_k$. We get
\begin{equation} 
  D_{\rm t} = {1+k_0^2 \over 2 \pi t_{\rm corr} k_0^2} \approx \{2.0,2.2,1.4,1.2\}\times 10^{-2}
\end{equation} 
for runs \{S,M,L,H\}. The L and H runs give reasonable values, but the S and M
runs have anomalously high mixing coefficients relative to other indicators of
the strength of turbulence. However, the random walk model overall appears
physically consistent and is useful in interpreting the simulations.

\subsection{Extrapolation to scales $\gg H$}

It is interesting to speculate how zonal flows would behave in even larger
boxes than considered here. This is relevant in discs with aspect ratio
$H/r\ll0.1$, such as the discs around supermassive black holes in active
galaxies \citep{Frank+etal2002,Goodman2003}.

Part of the observed growth in density amplitude with increasing box size is
the corresponding increase of the $c_k$ factor [eq.\ \ref{eq:rhoeq}]. Since
$c_k$ tends towards unity with increasing wavelength, it will not contribute to
the increase in $\hat{\rho}$ in boxes much larger than ten scale heights.
Assuming further that $\tau_{\rm for}$ and $\tau_{\rm mix}$ scale approximately
as $1/k_0^2$, and that $\hat{T}\propto k_0$, we predict that the density
amplitude will not continue to grow, but will remain relatively constant on
scales $\lambda\gg 10 H$.

However, this prediction is uncertain, given that we do not have a full
understanding of how the correlation time of the forcing depends on scale. Our
analysis in the previous section indeed showed that $\tau_{\rm for}$ grew
faster than $1/k_0^2$ from box L to box H. Also it is not known whether
$\widehat{B_x B_y}$ will remain constant at very large scales. Instead the
inverse cascade of magnetic energy may eventually terminate.

In larger boxes a wider range of scales can contribute to the density bumps,
because of the long mixing time-scale, not just the largest scale that we
considered above. This ``oligarchic'' regime may be already have been reached
in our run H ($L_x=10.56 H$) where the zonal flow amplitude is slightly smaller
than in run L, while the zonal flow peaks much less strongly at the largest
scale ($\zeta_{u_y}=-1.0$ for run L, but is only $\zeta_{u_y}=-0.2$ for run H).

\subsection{Towards a More Complete Theory}
\label{s:complete}

A more complete theory of large scale dynamics would include the
self-consistent evolution of magnetic fields. For instance, the magnetic
pressure grows from the Maxwell stress, via the term
\begin{equation}
  \frac{\dpa [(1/2)B_y^2]}{\dpa t} = \ldots - \frac{3}{2} \varOmega B_x B_y \, ,
\end{equation}
which readily follows from the induction equation (\ref{eq:eqindb}). Thus the
magnetic pressure  grows in regions of stronger (more negative) Maxwell stress.
The density, on the other hand decreases in regions of high stress.
\Fig{f:phasesum} clearly shows the anticorrelation of density (and thus thermal
pressure) with both Maxwell stresses and magnetic pressure. Though not shown in
the plot, actually all components of the magnetic pressure are found to be
lower in the density peak.

The departure from geostrophic balance for the imposed field runs M\_Bz\_*
likely arises from the relatively short correlation time of the more vigorous
turbulence in these simulations. This can break the validity of \Eq{eq:duxdteq}
by launching density waves. From \Eqs{eq:duydteq}{eq:dlnrhodteq} we see that
zonal flows can be launched immediately by the magnetic tension ($\de
\hat{u}_y/\de t=\hat{T}$), without leaking into the Coriolis force ($-\varOmega
u_x/2$) that gives radial compression. This way a pressure bump does not have
enough time to react to the zonal flow before the magnetic tension shifts
position.

The recent work by \cite{Vishniac2008} offers a model for the growth of large
scale structures in MRI turbulence, including an induction equation for field
evolution. Other features which differ from our model include incompressibility
and the neglect of azimuthal accelerations (while we ignored radial ones). Most
important, \cite{Vishniac2008} considers only disturbances with a sinusoidal
vertical dependence, which leads to the growth of vertically varying rolls. The
relation between these structures and the height-independent zonal flows that
emerge from our model deserves further study.

\section{Inverse Cascade}
\label{s:cascade}

So far we have shown that the large scale variation in the Maxwell stress
launches slightly compressive zonal flows, which in turn enter geostrophic
balance with axisymmetric column density bumps.

\begin{figure}
   \includegraphics[width=\linewidth]{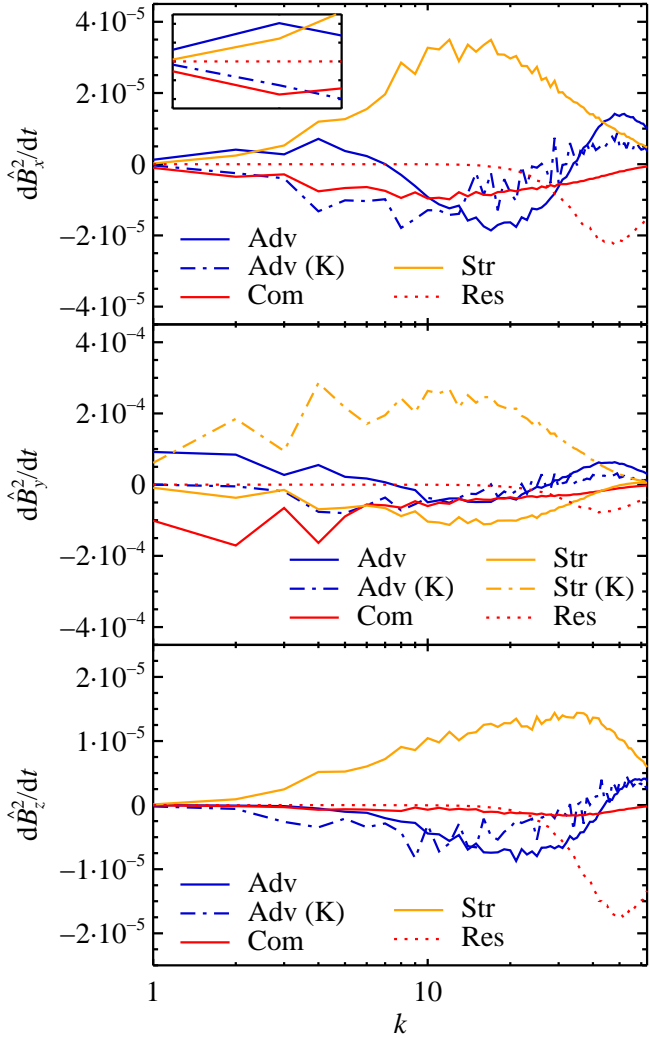}
   \caption{Contribution from different terms of the induction equation to the
   magnetic energy at scale $k$, summed over shells of constant $k=|\vc{k}|$,
   including the contribution from the hyperresistivity [``Res'']. Both the
   $x$- and $y$-components of the magnetic field get a positive energy
   injection at the largest scale from the advection term [``Adv'']. Another
   important energy source is the Keplerian stretching term [``Str (K)'']. The
   regular stretching term ([``Str'']) injects energy at all scales of $B_x$
   and $B_z$, but is an energy sink for $B_y$.}
  \label{f:energy_balance_bb_shell}
\end{figure}
\begin{figure}
   \includegraphics[width=\linewidth]{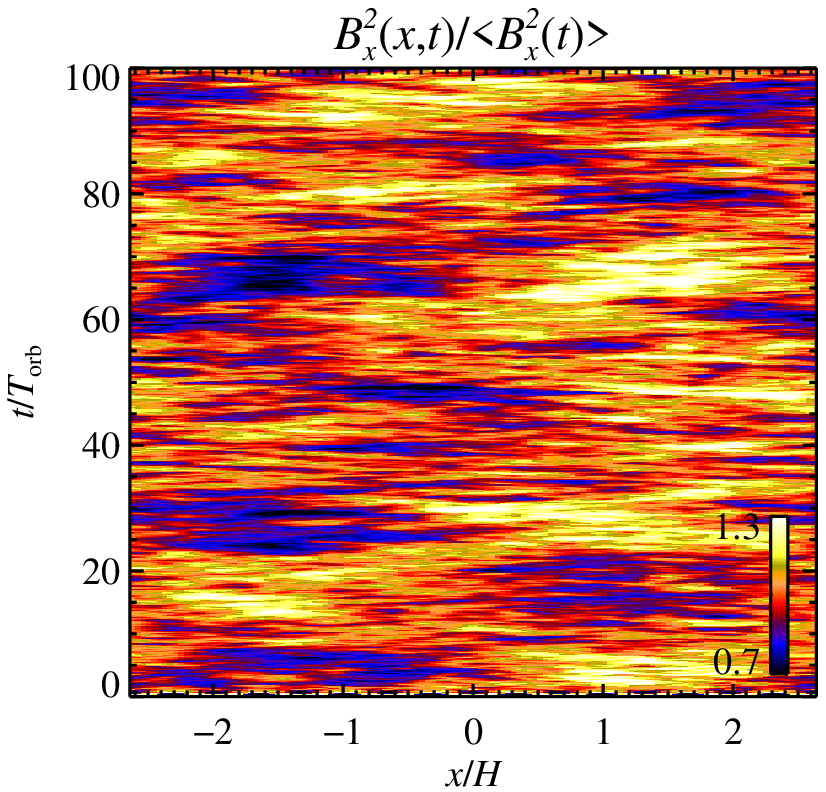}
   \includegraphics[width=\linewidth]{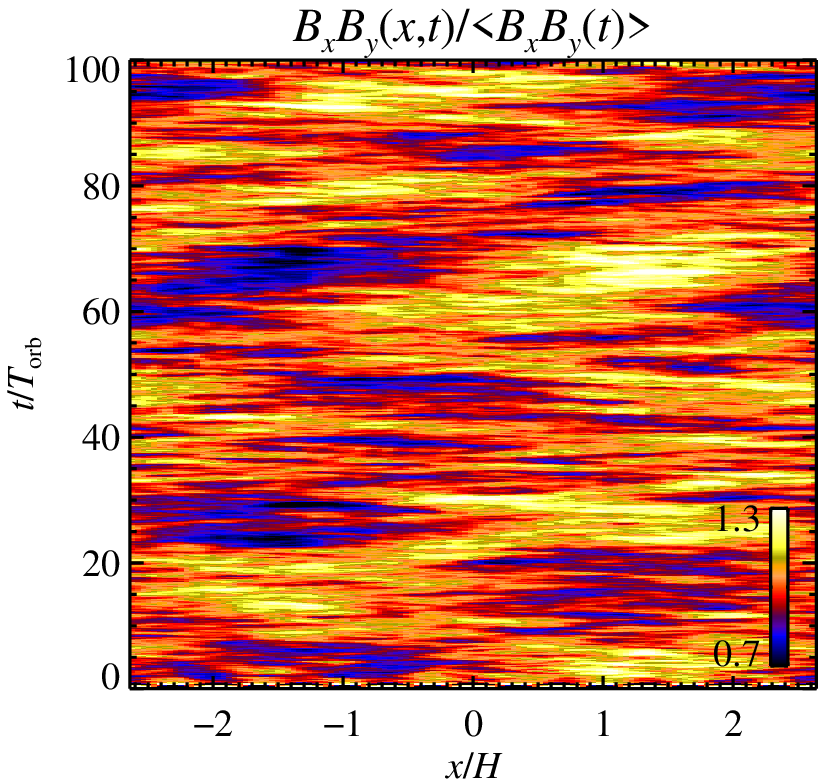}
   \caption{The radial magnetic energy (top panel) and the Maxwell stress
   (bottom panel) in run L, both averaged over the azimuthal and vertical
   directions and normalized by the instantaneous mean. There is an almost
   perfect correlation between the two quantities, arising from Keplerian
   stretching of the radial magnetic field. Energy and stress fluctuations peak
   at 30\% of the mean value, but the average fluctuation is only around 10\%.
   The coherence time is generally a few orbits, with periods of coherence
   times up to ten orbits. The correlation time of the density structures shown
   in \Fig{f:rhomx_t} is much longer.}
  \label{f:bxbymx_t}
\end{figure}
The magnetic energy in the azimuthal field is pumped to the largest radial
scales of the box by the advection term (in the induction equation, see
\Fig{f:energy_balance_all}). This inverse cascade is also evident from
\Fig{f:energy_balance_bb_shell}. Here we plot the power contribution from
different terms in the induction equation to the magnetic field, summing over
shells of constant wavenumber and averaging over equally spaced snapshots
during the saturated state of run L. The radial and azimuthal field both get a
net positive contribution from the advection term at large scales, whereas the
magnetic stretching term (including stretching by the background shear) adds
energy with a peak at middle scales around $k/k_0\sim20$. The source $B_x$ for
Keplerian stretching is itself generated by non-Keplerian stretching at both
moderate and small scales. For the vertical field component
\Fig{f:energy_balance_bb_shell} shows no sign of large scale magnetic energy
activity. This is interesting in connection with a possible accretion disk
dynamo, because the lack of large scale activity of $B_z$ may indicate that
vertical field, crucial for the vertical field MRI to operate, is only
replenished at scales near the dissipative subrange of the turbulence, but not
on the large scales. Azimuthal fields are unstable to a non-axisymmetric
magnetorotational swing instability \citep{BalbusHawley1992}, which has a
significant growth rate at short vertical wavelengths
\citep{FoglizzoTagger1995,TerquemPapaloizou1996}.
\Fig{f:energy_balance_bb_shell} indicates that the azimuthal field is actively
processing energy, through the various terms in the induction equation, even at
the largest scales of the box.

\Fig{f:energy_balance_bb_shell} also shows that compression is a sink of
magnetic energy at the largest scales, for both radial and azimuthal field. The
zonal flow launched by the radially dependent Maxwell stress indeed creates a
diverging velocity field in regions where the magnetic field is high, lowering
the magnetic energy at the largest scales. This way the zonal flows are
self-limiting, as they reduce the large scale variation in the magnetic field.

When the magnetic energy in the radial field component cascades up to the
largest scales of the box, then the Maxwell stress at those scales can grow
through the negative definite Keplerian stretching term,
\begin{equation}
  \dpa (B_x B_y)/\dpa t=\ldots - (3/2)\varOmega B_x^2 \, .
\end{equation} 
The deepest source of the zonal flows thus lies primarily in the inverse
cascade of the radial magnetic field energy. In \Fig{f:bxbymx_t} we show the
radial magnetic energy and the Maxwell stress in run L, averaged over the
azimuthal and vertical directions and normalized by the instantaneous mean.
There is an astonishing correlation between the two quantities.

\pagebreak
\subsection{Hydrodynamical Inverse Cascade?}

The slightly positive energy contribution of the advection term to the velocity
field (\Fig{f:energy_balance_all}) represents evidence for an additional
inverse cascade for the azimuthal velocity. Such an inverse cascade occurs
commonly in 2-D turbulence \citep{Kraichnan1967,Borue1993} and in 3-D rotating
turbulence \citep[e.g.][]{Bartello+etal1994,Cambon+etal1997}. In the latter
case vertically symmetric cyclones and anticyclones emerge at scales large
enough for Coriolis forces to dominate (i.e.\ Rossby number ${\rm Ro}\ll1$).
Cascade of small scale turbulent energy to large scale Rossby waves and zonal
jets in protoplanetary discs was proposed by \cite{Sheehan+etal1999}. However,
in our simulations the hydrodynamical inverse cascade is found to give a more
than 10 times smaller contribution to the large scale kinetic energy than the
variation in Maxwell stress and magnetic energy do.
\begin{deluxetable*}{lcccccccr}
  \tabletypesize{\small}
  \tablecaption{Zonal Flow Properties for Various Shearing Box Schemes}
  \tablewidth{0pt}
  \tablehead{
    \colhead{Run} &
    \colhead{$\langle -B_x B_y \rangle$} &
    \colhead{$|\hat{u}_y|(k_0,0,0)$} &
    \colhead{$|\hat{\rho}|(k_0,0,0)$} &
    \colhead{$|\widehat{B_x B_y}|(k_0,0,0)$} &
    \colhead{$\zeta_{u_y}$} &
    \colhead{$\zeta_{\rho}$} &
    \colhead{$\zeta_{B_x B_y}$} &
    \colhead{$t_{\rm corr}$}
    }
  \startdata
    S (FDA)    & $4.2\times10^{-3}$ & $1.9\times10^{-2}$ & $8.1\times10^{-3}$
               & $4.5\times10^{-4}$ & $-1.6$ & $-2.4$ & $-0.9$ &  $8.3$ \\
    S (RRC)    & $2.3\times10^{-3}$ & $1.8\times10^{-2}$ & $8.0\times10^{-3}$
               & $2.7\times10^{-4}$ & $-1.5$ & $-2.4$ & $-0.7$ & $14.9$ \\
    S (SRD)    & $3.1\times10^{-3}$ & $1.8\times10^{-2}$ & $7.9\times10^{-3}$
               & $3.8\times10^{-4}$ & $-1.5$ & $-2.3$ & $-0.8$ & $8.9$ \\
    S (SAFI)   & $2.8\times10^{-3}$ & $1.8\times10^{-2}$ & $8.1\times10^{-3}$
               & $3.7\times10^{-4}$ & $-1.5$ & $-2.5$ & $-0.8$ & $12.3$ \\
    \hline
    M (FDA)    & $1.2\times10^{-2}$ & $2.4\times10^{-2}$ & $1.9\times10^{-2}$
               & $1.2\times10^{-3}$ & $-1.1$ & $-1.8$ & $-0.7$ &  $8.5$ \\
    M (RRC)    & $8.6\times10^{-3}$ & $2.6\times10^{-2}$ & $2.4\times10^{-2}$
               & $9.7\times10^{-4}$ & $-1.3$ & $-2.3$ & $-1.0$ & $11.6$ \\
    M (SRD)    & $8.8\times10^{-3}$ & $1.7\times10^{-2}$ & $1.7\times10^{-2}$
               & $8.7\times10^{-4}$ & $-0.5$ & $-1.8$ & $-0.8$ &  $8.3$ \\
    M (SAFI)   & $9.6\times10^{-3}$ & $1.9\times10^{-2}$ & $1.6\times10^{-2}$
               & $9.2\times10^{-4}$ & $-0.8$ & $-1.8$ & $-0.7$ &  $8.4$ \\
    \hline
    L (FDA)    & $1.1\times10^{-2}$ & $2.5\times10^{-2}$ & $5.0\times10^{-2}$
               & $1.3\times10^{-3}$ & $-1.1$ & $-2.3$ & $-1.0$ & $18.6$ \\
    L (RRC)    & $1.3\times10^{-2}$ & $1.8\times10^{-2}$ & $3.6\times10^{-2}$
               & $1.1\times10^{-3}$ & $-0.5$ & $-1.6$ & $-0.5$ & $20.1$ \\
    L (SRD)    & $1.1\times10^{-2}$ & $2.0\times10^{-2}$ & $4.0\times10^{-2}$
               & $1.2\times10^{-3}$ & $-0.6$ & $-1.9$ & $-0.9$ & $20.9$ \\
    L (SAFI)   & $1.1\times10^{-2}$ & $1.7\times10^{-2}$ & $3.3\times10^{-2}$
               & $1.3\times10^{-3}$ & $-1.0$ & $-2.3$ & $-1.4$ & $19.6$ \\
    \hline
    H (FDA)    & $9.5\times10^{-3}$ & $1.6\times10^{-2}$ & $5.8\times10^{-2}$
               & $1.1\times10^{-3}$ & $-0.4$ & $-1.4$ & $-1.1$ & $57.1$ \\
    H (SAFI)   & $1.0\times10^{-2}$ & $1.5\times10^{-2}$ & $5.2\times10^{-2}$
               & $1.1\times10^{-3}$ & $-0.2$ & $-1.1$ & $-0.9$ & $50.5$ \\
  \enddata
  \tablecomments{Col.\ (1): Name of run. Col.\ (2): Maxwell stress, normalized
  by mean pressure. Col.\ (3)-(5): Fourier amplitude of azimuthal velocity,
  density normalized by mean density in the box, and pressure-normalized
  Maxwell stress, respectively, at largest radial mode. Col.\ (6)-(8): Power
  law index of azimuthal velocity, density, and Maxwell stress, respectively,
  calculated from two largest radial modes. Col.\ (9): Correlation time, in
  orbits $T=2\pi\varOmega^{-1}$, of largest radial density mode. The various
  shearing box algorithms -- FDA, RRC, SRD, and SAFI -- are defined in
  \App{s:xdepdiff}.}
  \label{t:zonalflow2}
\end{deluxetable*}

\section{Numerical Issues and Validation Tests}
\label{s:validation}

Supersonic shear flow in large shearing boxes introduces a space-dependent
numerical dissipation which can lead to artificial suppression of turbulence
near the edges of the box \citep{Johnson+etal2008}. In this section we present
various tests to validate the results of the non-linear simulations.

\subsection{Results for Various Shearing Box Algorithms}

In \App{s:xdepdiff} we show variations of runs S--H where we use several
different ways to solve the shearing box equations. The various schemes -- FDA
(Finite Difference Advection), RRC (Random Revolution Center), SRD (Systematic
Radial Displacement), and SAFI (Shear Advection by Fourier Interpolation) --
are discussed in detail in \App{s:xdepdiff}. The statistical properties of the
zonal flows and the pressure bumps are given in \Tab{t:zonalflow2}. The large
scale variation in the Maxwell stress remains at around 10\% of the mean
Maxwell stress, independent of the shearing box algorithm. The amplitudes of
azimuthal velocity and density, on the other hand, are clearly larger when
applying the straightforward finite difference advection scheme, indicating
that this scheme artificially enhances the correlation time of the Maxwell
stress. For the standard finite difference advection scheme of the Pencil Code
we furthermore observe a tendency for turbulence to be weaker near the edge of
the box and for the largest radial density mode to have its minimum near the
center of the box (see \Fig{f:bxbymx} of \App{s:xdepdiff}) for the two largest
boxes, confirming the results of \cite{Johnson+etal2008}. However, with the two
improved shearing box algorithms Systematic Radial Displacement (SRD) and Shear
Advection by Fourier Advection (SAFI), any consistent preference for stronger
turbulence near the box center is no longer present. This was our primary
reason for presenting the SAFI versions of L and H in the main text.

\subsection{Convergence Tests}

To quantify the effect of dissipation parameters and grid resolution on the
amplitude of zonal flows and on the inverse cascade of magnetic energy we run
two sets of simulations: (a) runs M\_Bz at three different resolutions with
decreasing diffusivity coefficients and (b) runs MS at two different
resolutions, the higher resolution simulations run both with a fixed and with a
decreased diffusivity. The results are given in \Tab{t:turbulence} and
\Tab{t:zonalflow}.

For the imposed field simulations M\_Bz there is an overall increase in the
turbulent activity of 50\% when going from the lowest (M\_Bz\_r0.5) to the
highest resolution (M\_Bz\_r2), while the zonal flow and the density bumps
decrease by around 20\% in amplitude. In this series of runs we have decreased
the hyperdiffusivity coefficients proportional to $\delta x^5$ with increasing
resolution. The increase in turbulent stresses may thus be due to the ever
larger inertial range. At the same time these added scales will act as an
increased diffusivity on the large scale pressure bumps, so it is not
surprising that the zonal flows fall slightly in amplitude.

The MS series of runs are non-stratified and have a computational domain of
$L_x=L_y=2.64 H$ and $L_z=1.32 H$. For the low resolution simulation (MS) we
have forced a time-step similar to that of the higher resolution run
(MS\_r2\_fix), to have approximately the same amount of numerical dissipation,
and we apply the Systematic Radial Displacement scheme described in
\App{s:xdepdiff} to ensure that the numerical dissipation does not have
significant dependence on $x$. We find that zonal flows decrease in amplitude
when increasing the resolution while keeping the {\it mesh} hyper-Reynolds
number constant (compare run MS to run MS\_r2 in \Tab{t:zonalflow}). The mesh
hyper-Reynolds number is here defined as ${\rm Re}_3 = {\rm max}(u) \delta
x^5/\nu_3$. However, the overall turbulent activity also falls significantly as
the resolution is increased, so the reduction in the large scale amplitudes may
simply be due to the decrease in available power at the small scales. On the
other hand if dissipation parameters are kept constant as the resolution is
increased (as in run MS\_r2\_fix), then there is little or no effect on
turbulent activity and on zonal flow/pressure bump statistics. It is
encouraging that extra spatial resolution has so little effect on the turbulent
state when the dissipation, and thus spectral range, is held fixed. Ultimately
even higher resolution simulations will be needed to fully discern the effect
of scale separation and Reynolds numbers on the zonal flow amplitude.

\subsection{Laplacian Dissipation}

The simulations we have presented so far all used hyperdiffusivity operators to
dissipate energy near the smallest scales of the simulation. This is a
``trick'' that allows us to extend the inertial range of the turbulence, at the
cost of a steeper transition into the dissipative subrange. To check that the
use of hyperdiffusivity does not create any spurious flows, we have run three
simulations using regular Laplacian operators for viscosity and resistivity
(runs SS\_r4\_nu1, SS\_r8\_nu1, and MS\_r4\_nu1). In order to ensure that the
box can sustain turbulence, we set the magnetic Prandtl number (defined as
ratio of viscosity coefficient $\nu_1$ to resistivity coefficient $\eta_1$) to
approximately ${\rm Pm}\equiv \nu_1/\eta_1=4$ \citep[following][who found that
turbulence in zero net flux boxes without stratification is generally only
sustained when ${\rm Pm}>1$]{Fromang+etal2007}. Still the Pencil Code only
showed sustained turbulence at a resolution of at least $100$ grid points per
scale height, as lower resolution would either show decaying turbulence at low
magnetic Reynolds numbers, or crash due to lack of dissipation at high magnetic
Reynolds numbers.

For moderate and high resolutions, the results in \Tab{t:turbulence} and
\Tab{t:zonalflow} show that the Laplacian simulations exhibited fairly strong
turbulence with just as strong zonal flows as the hyperdiffusivity runs. This
yields much support to the fact that the zonal flows are real, and not a
numerical artifact, because of the more physical nature of the Laplacian
operators and the fact that zonal flows operate also at $200$ grid points per
scale height in resolution, a factor eight times higher resolution than the
hyperdiffusivity runs S--H. However there is a significant increase in
turbulent activity when going from run SS\_r4 to run SS\_r8. The fact that the
MRI has not yet converged even at such high resolution is a slight worry. It
may indicate either that numerical dissipation still influences the saturation
level of the turbulence or that the introduction of smaller scales changes the
dynamics of the system. One would hope that the MRI will converge at the next
resolution level, i.e.\ 400 grid points per scale height, but this huge
computational effort is beyond the scope of the current investigation. The
ratio of the large scale Maxwell stress variation to the mean value
nevertheless stays approximately the same for the two runs SS\_r4 and SS\_r8,
indicating some convergence in the inverse cascade of magnetic energy.

\section{Summary and Discussion}
\label{s:discussion}

We conduct numerical simulations of turbulence driven by the magnetorotational
instability (MRI) in large shearing boxes. We here emphasize four main
results. (1) The turbulent energy and stresses more than double when the box
size (radial and azimuthal) increases from $1.32$ to $2.64$ scale heights. In
yet larger boxes the turbulent fluctuations and stresses remain relatively
constant. (2) Magnetic energy exceeds kinetic energy by a factor two-three on
scales above approximately two scale heights, when the box size is sufficient
for the inverse case to proceed to these large scales. (3) Axisymmetric surface
density fluctuations grow to fill the box, and persist for many orbits, with an
increasing lifetime in bigger boxes. These pressure bumps are in geostrophic
balance with sub/super-Keplerian zonal flows. (4) We put forward a simple model
for the launching of zonal flows by large scale variations in the Maxwell
stress. This random walk model explains both the saturated amplitude as well as
the correlation time of pressure bumps and their associated zonal flows.

Large scale pressure and zonal flow structures grow to fill the simulation
domain for all box sizes. The largest box size considered, more than ten scale
heights in radial and azimuthal extent, nevertheless indicates an increasingly
flat dependence of zonal flow power on the radial wave number. Even larger
simulation domains will be needed to verify whether the natural scale of the
large scale structures is really around ten scale heights. Global effects --
including curvature and radial variation in background quantities like density
and sound speed -- may also play a vital role for the termination of the
inverse cascade. The global simulations by \cite{Lyra+etal2008} show evidence
of long lived radial pressure bumps that have not cascaded to the largest
radial scale of the disk (see their Fig.\ 18). It is intriguing to note that
the length scale of the pressure bumps in \cite{Lyra+etal2008} is around ten
scale heights.

We have experimented with several improvements of the shearing box algorithm in
order to quantify the dependence of the results on numerical diffusivity
associated with the Keplerian shear flow. There is a tendency for the density
depression to occur at the center of the box for the two largest box sizes
(approximately five and ten scale heights in radial extent, respectively) if
shear advection is done by the usual finite difference scheme of the Pencil
Code, confirming a similar effect found by \cite{Johnson+etal2008} for a
ZEUS-like code \citep{StoneNorman1992}. However, when shear advection is
integrated by high order interpolation instead, we see no preference for any
special location in the box. The same is true if the state of the gas is
displaced systematically by one grid point in the radial direction at the
beginning of each time-step.

Ultimately the emergence of zonal flows and pressure bumps is controlled by the
large scale variation of the magnetic field. The cascade of magnetic energy
from small to large scales only occurs for the $x$- and $y$-components of the
field. The vertical field, which is of importance for feeding the vertical
field MRI, shows very little large scale behavior. The specific mechanism that
leads to the inverse cascade of the in-plane field in our simulations is not
known. In the absence of any net kinetic helicity separation in the radial
direction, the large scale field may arise as an effect of a negative turbulent
diffusivity \citep[e.g.][]{Zheligovsky+etal2001,Urpin2002} or a shear current
effect \citep{RogachevskiiKleeorin2003,Yousef+etal2008}.

Both hyperdiffusivity simulations and direct numerical simulations (DNS), with
Laplacian viscosity and resistivity operators, show prodigious zonal flows. A
fruitful future path could be to focus on even higher resolution direct
simulations of resistive MRI turbulence. This could probe the dependence of the
inverse cascade of magnetic energy and the launching of zonal flows on the
magnetic Reynolds number and the Prandtl number
\citep{LesurLongaretti2007,Fromang+etal2007}, believed to be much smaller than
unity in most parts of accretion disks \citep{BalbusHenri2008}.

Our numerical experiments show that the magnetic field takes particular
advantage of larger spatial scales by participating in an inverse cascade. This
in turn produces long-lived density fluctuations which can have significant
influence on the formation and evolution of planetary systems. Big shearing
boxes thus provide an excellent tool for studying turbulent accretion disks and
the physical processes that occur therein.

\acknowledgments

We are grateful to Takayoshi Sano, Ethan Vishniac, Axel Brandenburg, Eric
Blackman, Yuri Levin and Wladimir Lyra for inspiring discussions on large scale
magnetic fields in accretion disks. We would like to thank the anonymous
referee for an insightful referee report. Bryan Johnson is thanked for
commenting an early version of the manuscript. H.~K. has been supported in part
by the Deutsche Forschungsgemeinschaft DFG through grant DFG Forschergruppe 759
``The Formation of Planets: The Critical First Growth Phase''.

\appendix

\section{Test of Shearing Box Implementation}\label{s:shearwave}

In this Appendix we test the shearing box implementation of the Pencil Code
against known analytical solutions to the evolution of shearing waves. In
\S\ref{s:hshearwave} we compare the results of the Pencil Code to the
analytical solution found by \cite{BalbusHawley2006} for hydrodynamical
shearing waves. We proceed to test the magnetic field module in
\S\ref{s:mshearwave}, where we compare a numerical solution of the linearized
shearing box MHD equations to the results obtained with the Pencil Code. We
find excellent agreement in both cases.

\subsection{Time-Dependent Incompressible Shear Wave}
\label{s:hshearwave}

\cite{BalbusHawley2006} derived a time-dependent, non-axisymmetric and
incompressible solution for hydrodynamical shear waves. Solutions are of the
form of a single mode $f(x,y,z,t)=\hat{f}(t)\exp[\ii(k_x(t) x+k_y y+k_z z)]$
for the dynamical variables $u_x$, $u_y$ and $u_z$. The Keplerian shear causes
the radial wavenumber to be time-dependent and have the form
$k_x(t)=k_x(0)+(3/2)\varOmega t k_y$.

We have performed numerical simulations of the model set up described in \S5 of
\cite{BalbusHawley2006}. We use the simulation parameters $q=3/2$,
$\varOmega=0.001$, $\rho=1$, $c_{\rm s}=4.08\varOmega$, $\gamma=5/3$ and
$P=10^{-5}$. The box size is $L_x=L_y=L_z=1$. The initial condition is a
non-axisymmetric wave in the $x$-velocity,
\begin{equation}
  u_x(x,y,z,t=0)=A_R \sin[2\pi(y+4z)] \, ,
\end{equation}
with the amplitude $A_R=10^{-5}$.

\begin{figure*}
   \begin{center}
     \includegraphics[width=8.7cm]{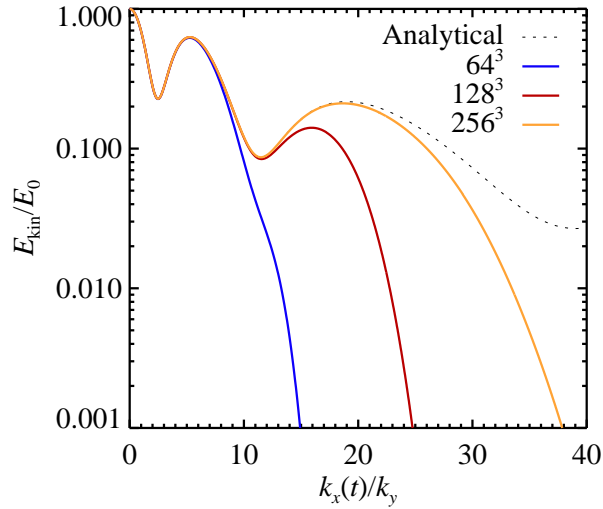}
   \end{center}
   \caption{Normalized kinetic energy of a time-dependent non-axisymmetric
   shear wave, for different numerical resolutions and the analytical
   solution. The numerical solution follows the analytical solution well down
   to around 6-8 grid points per radial wavelength, after which the explicit
   viscosity is so strong that the shear wave dies out. There is no evidence of
   non-linear aliasing as the wave approaches the grid scale, again due to
   damping by the explicit viscosity terms at the small scales. Compare with
   Fig.\ 3 of \cite{BalbusHawley2006} which shows the results with the Athena
   code.}
  \label{f:shearwave}
\end{figure*}

In \Fig{f:shearwave} we show the numerical solution for a range of resolutions
and compare with the analytical solution. There is excellent agreement between
the Pencil Code integration and the analytical solution, down to 6-8 grid
points per wavelength where the viscous damping is so strong that the wave dies
out. There is no trace of non-linear aliasing as the wave is sheared below the
spatial resolution. \cite{BalbusHawley2006} show a similar plot for the Athena
code in their Fig.\ 3. Actually the Pencil Code is slightly less dissipative
than the Athena code, but this of course depends on our choice of dissipation
parameters. For the results in \Fig{f:shearwave} we set the hyperviscosity and
hyperdiffusion similar to their values in the MRI simulations (i.e.\ similar
viscous dissipation time at the grid scale).

\subsection{Magnetized Shear Wave in the Linear Regime}
\label{s:mshearwave}

To test the evolution of the magnetic vector potential in the shearing box we
set up a magnetized shearing wave test problem. Magnetized shear waves are an
interesting analytical and numerical topic, but in this appendix we shall
consider such waves solely to test the numerical solver of the Pencil Code. We
refer to \cite{Johnson2007} for details on the physical behavior of magnetized
shear waves.
\begin{figure*}
   \includegraphics[width=\linewidth]{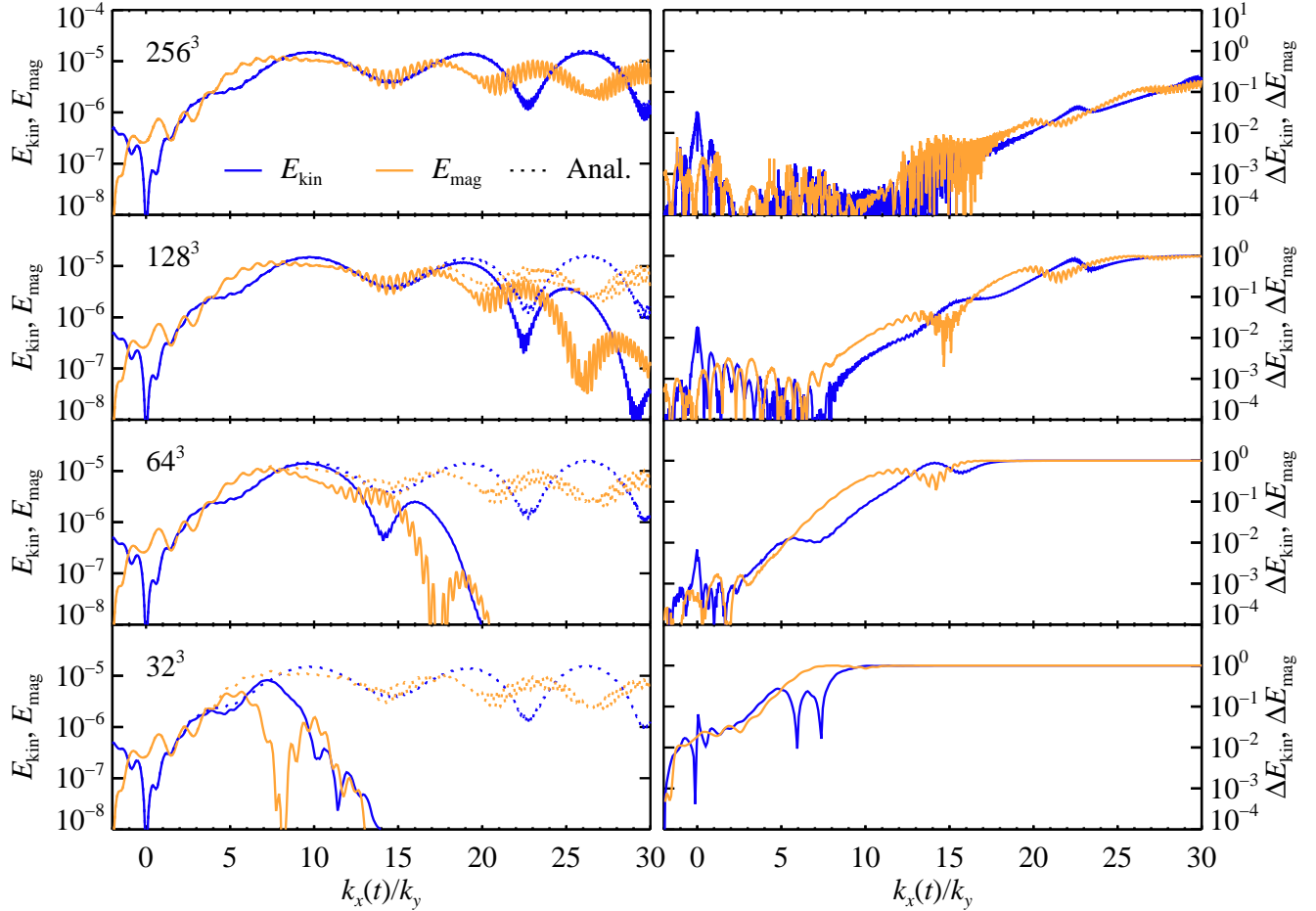}
   \caption{Kinetic and magnetic energy of a magnetized, non-axisymmetric shear
   wave as a function of radial wavenumber (left plots). The relative error
   between analytical and numerical solution is shown in the right plots (here
   zero is defined as no error, while an error measure of unity corresponds to
   infinite error). There is excellent agreement between the numerical solution
   of the Pencil Code and the analytical solution, down to approximately 6 grid
   points per wavelength.}
  \label{f:mshearwave}
\end{figure*}

Linearizing the equation of motion, continuity equation and induction equation
as $f=f_0 + f'$ for all components and considering the time evolution of
individual Fourier modes of the form
\begin{equation}
  f'(x,y,z,t) = \hat{f}(t) \exp[\ii(k_x(t) x + k_y y + k_z z)] \, ,
\end{equation}
with time-dependent radial wavenumber
\begin{equation}
  k_x(t) = k_x(0)+\frac{3}{2}\varOmega t k_y \, ,
\end{equation}
yields
\begin{eqnarray}
  \frac{\de \hat{\vc{u}}}{\de t} &=&
    2 \varOmega  \hat{u}_y \vc{e}_x - \frac{1}{2} \varOmega \hat{u}_x \vc{e}_y
    + \frac{1}{\rho_0} \left[ -\ii\vc{k}(\vc{B}_0\cdot\hat{\vc{B}})
    +\ii(\vc{B}_0\cdot\vc{k})\hat{\vc{B}}\right]
    - \frac{c_{\rm s}^2}{\rho_0} \ii \vc{k} \hat{\rho} \, , \\
  \frac{\de \hat{\rho}}{\de t} &=& -\rho_0 \ii \vc{k}\cdot\hat{\vc{u}} \, , \\
  \frac{\de \hat{\vc{B}}}{\de t} &=&
    \ii (\vc{B}_0\cdot\vc{k})\hat{\vc{u}}
    - \frac{3}{2} \varOmega \hat{B_x} \vc{e}_y -
    \ii \vc{B}_0\vc{k}\cdot\hat{\vc{u}} \, .
\end{eqnarray}
Setting $\vc{B}=(0,0,B_0)$, $\vc{k}=(0,0,k_z)$,
$\hat{\vc{u}}=(\hat{u}_x,\hat{u}_y,0)$ and
$\hat{\vc{B}}=(\hat{B}_x,\hat{B_y},0)$ would yield evolution according to the
linear magnetorotational instability \citep{BalbusHawley1991}. But since we are
interested in the potential effect of shear advection and shear-periodic
boundary conditions we instead seed a compressive, non-axisymmetric mode with
$\vc{k}=[-2,1,4]$, $\hat{\vc{u}}=[0.001,0,0]$, $\rho_0=1$, $\vc{B}_0=[0,1,0]$,
$\hat{\vc{B}}=[0,0,0]$ in a box of size $(2\pi)^3$. Such a leading wave
perturbation of an azimuthal field line is subject to a transient
non-axisymmetric magnetorotational instability
\citep{BalbusHawley1992,FoglizzoTagger1995,TerquemPapaloizou1996}. In
\Fig{f:mshearwave} we compare the evolution of magnetic and kinetic energy
obtained by the Pencil Code to the evolution obtained by integrating the
linearized equation system in time. There is excellent agreement between the
two down to approximately 6 grid points per wavelength.

\section{Keplerian Shear Advection as Interpolation}
\label{s:safi}

In this Appendix we describe the numerical implementation of interpolated
Keplerian advection in the Pencil Code. The splitting of the advection step
into advection by Keplerian shear and advection by the fluctuating velocity
goes back to \cite{Gammie2001} for the shearing box approximation and to
\cite{Masset2000} for global, cylindrical coordinates. The interpolation scheme
was recently generalized to include magnetic fields, in the shearing box
approximation, and implemented in a ZEUS-like code by \cite{Johnson+etal2008}.

All shearing box dynamical variables $f$ have a Keplerian advection term of the
form
\begin{equation}
  \frac{\dpa f}{\dpa t} = -u_y^{(0)}\frac{\dpa f}{\dpa y}
\end{equation}
in their time evolution equation. Here $u_y^{(0)}=-(3/2)\varOmega x$ is the
linearized Keplerian velocity. The time-step associated with the Keplerian
advection is $\delta t = c_{\delta t} \delta x/{\rm max}(|u_y^{(0)}|)$, where
$c_{\delta t}$ is a factor of less than unity (in the Pencil Code stability
requires approximately $c_{\delta t}\leq0.4$). The Keplerian advection
time-step dominates over the time-step from sound waves when ${\rm
max}(|u_y^{(0)}|)>c_{\rm s}$, i.e.\ when $L_x/H>2/(3/2)\approx1.33$ for
Keplerian rotation.

The time-step of the Keplerian advection is irrelevant if that term is not
treated by finite-differencing, but rather as a shift of all variables in
physical space by the amount $u_y^{(0)} \delta t$. Without deriving it formally
we can estimate that the stability criterion with this method is that
neighboring points in the $x$-direction must not move more than a fraction
$\xi$ of a grid point apart,
\begin{equation}
  (3/2) \varOmega \delta x \delta t < \xi \delta y \, .
\end{equation}
This is a much less tight time-step constraint than for the finite difference
approach, since there is no longer any reference to the radial coordinate $x$
in the time-step calculation. The interpolation scheme also eliminates the
numerical diffusivity associated with the Keplerian advection (see
\App{s:xdepdiff}).

We implement the interpolated Keplerian advection as follows:
\begin{enumerate}
  \item All variables $q(x,y,z)$ are Fourier transformed in the $y$-direction
  to yield $\hat{q}(x,k_y,z)$.
  \item Each Fourier amplitude is multiplied by the complex factor $\exp[\ii
  k_y u_y^{(0)}(x) \delta t]$ to shift by the amount $u_y^{(0)}(x) \delta t$ in
  real space.
  \item Inverse Fourier transform to real space.
\end{enumerate}
Because of the third order Runge-Kutta time-integration scheme of the Pencil
Code, the shift must be done for the inherited right-hand-sides $\dot{q}$ at
the beginning of the second and third sub-time-steps as well. The time-step
entering in point 2 above, now dominated by sound waves, is calculated at the
first sub-time-step of the Runge Kutta scheme and consequently used for all
three substeps, including the Keplerian advection.

Note that in \cite{Masset2000} the state is always shifted by an integer number
of grid points, treating the remaining Keplerian advection speed together with
the perturbed speed. In this paper, on the other hand, the entire Keplerian
advection is done by interpolation in Fourier space. We denote this scheme
Shear Advection by Fourier Interpolation (SAFI).

\section{Removing Position-Dependent Numerical Diffusivity}
\label{s:xdepdiff}

In this Appendix we quantify the numerical diffusivity of the Pencil Code
advection scheme and, very importantly for simulations of turbulence in large
shearing boxes, its dependence on the background advection speed. We show how
to eliminate the spatial dependence of the numerical diffusivity by various
improvements to the shearing box algorithm of the Pencil Code.

\subsection{Measuring the Numerical Diffusivity of the Pencil Code}

Although numerical diffusion is not necessarily representable as a well-behaved
differential operator, we shall fit the time evolution of a passively advected
physical variable $f$ according to the expression
\begin{equation}
  |\hat{f}|(t) = |\hat{f}|(0) \exp(-D_n k^{2 n} t) \, .
\end{equation}
The hat denotes here that $f$ is set as a pure Fourier mode at scale $k$, and
$D_n$ is a diffusivity coefficient of order $n$. Regular diffusion would e.g.\
have $n=1$, while hyperdiffusion has higher values of $n$.

\begin{figure*}
   \includegraphics[width=\linewidth]{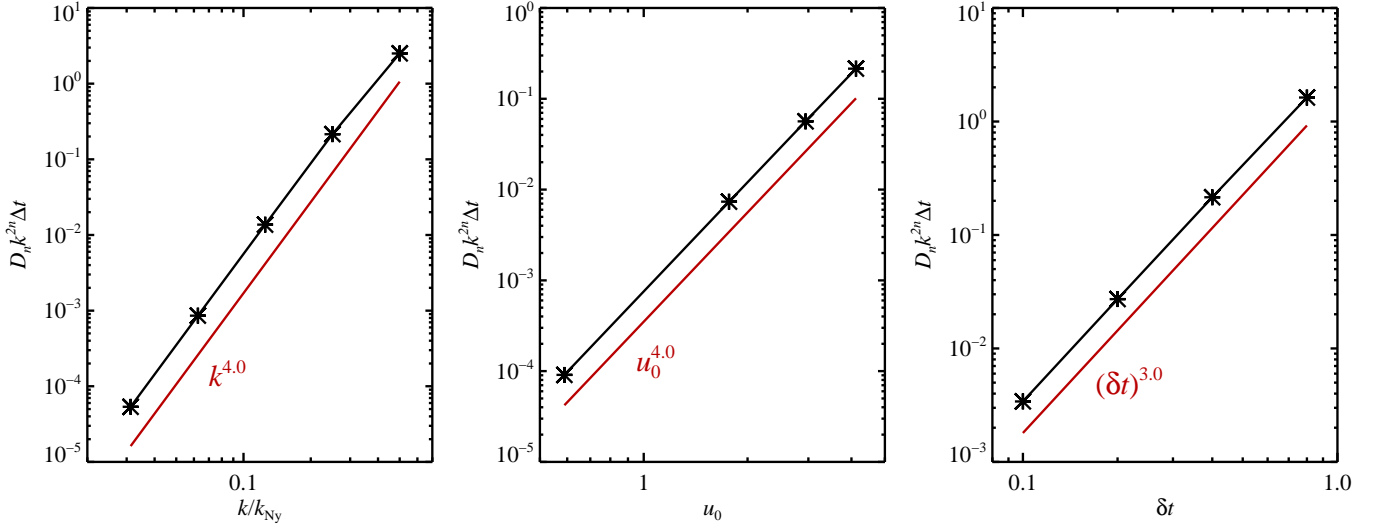}
   \caption{The dependence of the amplitude reduction parameter $Q=D_n k^{2n}
   \Delta t$ on wavenumber $k$ (first plot), advection speed $u_0$ (middle
   plot) and time-step $\delta t$ (last plot). All dependencies are fitted
   excellently by a power-law, indicating an numerical diffusivity that behaves
   like second order hyperdiffusivity.}
  \label{f:amplitudered}
\end{figure*}

We measure the diffusivity of the Pencil Code's advection scheme by advecting a
wave of arbitrary initial amplitude for the time $\Delta t=100$ with the
constant advection speed $u_0$. We then fit the amplitude reduction parameter,
\begin{equation}
  Q = -{\rm ln}\left( \frac{|\hat{f}|(\Delta t)}{|\hat{f}|(0)} \right) \, ,
\end{equation}
according to the power law
\begin{equation}
  Q \propto k^{2 n} u_0^m (\delta t)^l \, .
\end{equation}
Here $\delta t$ is the time-step of the temporal integration. If the
wavenumber-dependence can indeed be fitted with a single $n$, then the
diffusion coefficient is fitted according to $D_n\propto u_0^m (\delta t)^l$.
We show in \Fig{f:amplitudered} the dependence of $Q$ on $k$, $u_0$ and $\delta
t$. There is an extremely good power-law fit for each parameter, with $n=2$,
$m=4$ and $l=3$. Thus the numerical diffusivity behaves like a second order
hyperdiffusion,
\begin{equation}
  \dot{f} = - D_2 \frac{\dpa^4f}{\dpa y^4} \, ,
\end{equation}
with diffusion coefficient
\begin{equation}
  D_2 \approx 0.041 u_0^4 (\delta t)^3 \, .
\end{equation}
One may in fact show from stability analysis that $D_2 = (1/24) u_0^4 (\delta
t)^3$ (W.\ Dobler, private communication). The strong dependence on the
time-step $\delta t$ is not surprising for the 3rd order Runge-Kutta scheme of
the Pencil Code. One consequence is that if $\delta t$ is determined from the
global maximum speed, and $u_0$ varies across the simulation domain, then the
diffusion coefficient goes as the background speed to the fourth power, which
can lead to huge differences in the numerical diffusivity across the box.
\begin{figure*}
   \includegraphics[width=\linewidth]{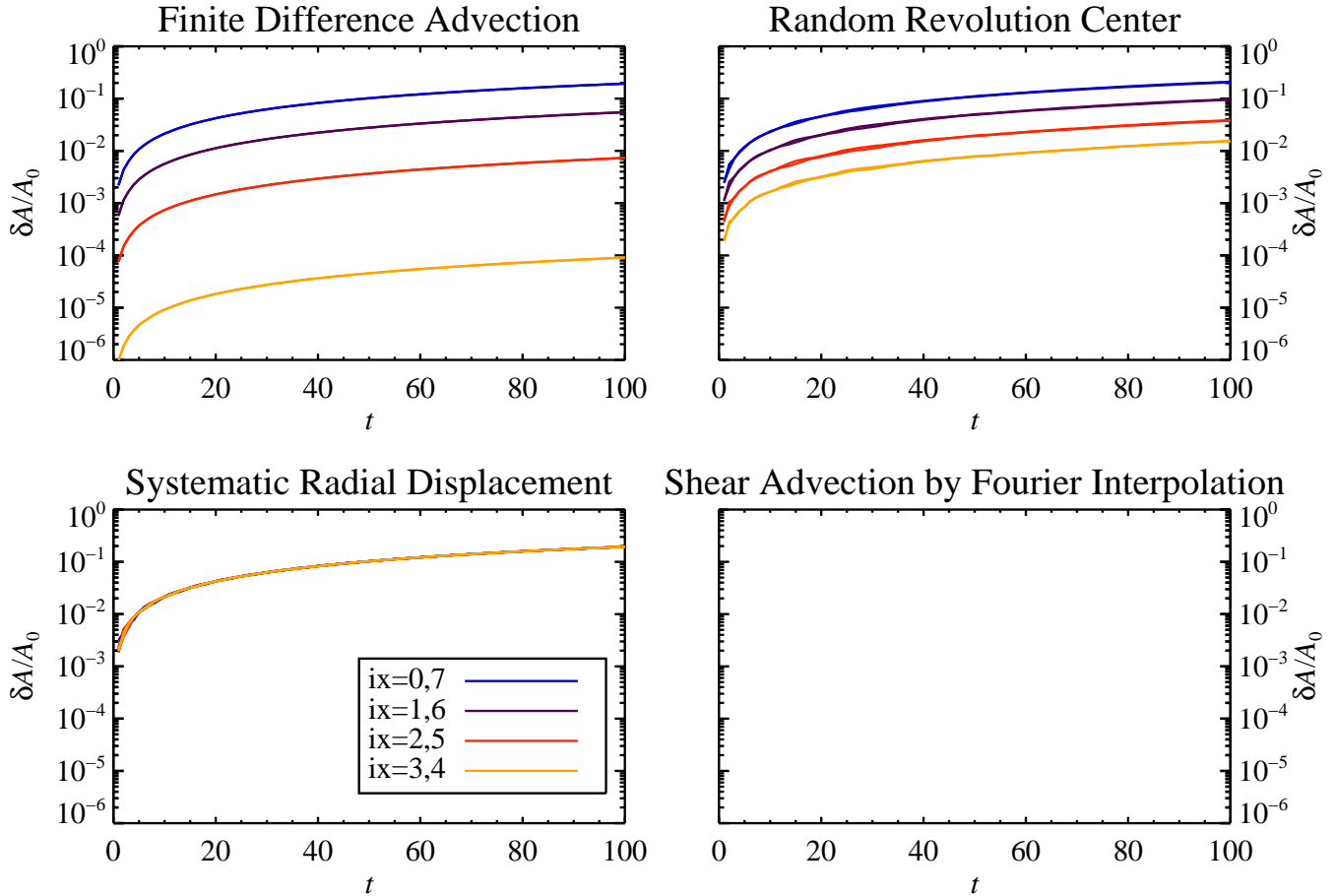}
   \caption{The amplitude of a passively advected shear wave as a function of
   time $t$ at four radial positions in the box (different colors). The
   amplitude falls due to the numerical diffusivity associated with the
   Keplerian advection term. The simple finite difference scheme gives much
   higher error near the edges of the box (first plot), as the numerical
   viscosity scales with the local advection speed to the fourth power.
   Centering the shear around a random point in the box increases the error in
   the center (second plot), but the error maintains a significant dependence
   on the radial coordinate. By displacing the entire data cube by one grid
   point in the radial direction in the beginning of each time-step or by
   interpolating the shear advection term in Fourier space one can remove the
   radial dependence of the amplitude error (third and fourth plots).}
  \label{f:xdependent_error}
\end{figure*}

\subsection{Improved Shearing Box Algorithms}

We proceed to test the effect of the varying numerical diffusivity in a linear
shear flow. We have set up a simple advection test where a wave of azimuthal
wavelength 8 grid points is advected by a linear shear flow with velocity
difference $(3/2)\pi\approx4.71$ from box edge to box center (the radial extent
of the box is $L_x=2\pi$). We follow the amplitude of the wave as a function of
the distance from the center of the box. This is plotted in
\Fig{f:xdependent_error}. The first panel shows that for the usual advection
scheme the error grows much quicker near the box edges, while the box center is
left almost untouched by numerical diffusion\footnote{The central grid points
$i=3,4$ lie at both sides of $x=0$, thus even these grid points experience some
shear advection.}.

There are several ways to reduce, or even remove, this space-dependent
numerical diffusivity. Here we discuss three ways:
\begin{enumerate}
  \item Picking the center of revolution randomly at each time-step, thus
  making sure that even the central points are advected by the shear.
  \item Displacing the entire data cube by one grid point in the radial
  direction in the beginning of each time-step.
  \item Treating the shear advection term separately by high order
  interpolation.
\end{enumerate}

\subsubsection{Random Revolution Center (RRC)}

The linear shear flow can be generalized as
\begin{equation}
  u_y^{(0)} = -\frac{3}{2} \varOmega (x-x_0) \, ,
\end{equation}
where $x_0$ is the radial coordinate of the revolution center, normally set to
zero. We can nevertheless choose an arbitrary $x_0\in[-L_x/2,L_x/2]$ at each
time-step and still solve exactly the same equation system. This way we can
reduce the special state that $x=0$ normally has because that point is not
advected. The second plot of \Fig{f:xdependent_error} shows the radially
dependent amplitude error using the Random Revolution Center scheme. While the
error near the center is greatly increased compared to the usual advection
scheme, there is still a significant difference in the amplitude between the
center and the edges of the box.

The time-step constraint of the RRC method is slightly tighter than for the
usual advection scheme, since the maximum advection speed is around a factor
two times higher. The higher average advection speed leads to some increase in
the overall numerical diffusivity, although the lower time-step partially
counteracts this increase.

\subsubsection{Systematic Radial Displacement (SRD)}

A better way to reduce the radially dependent numerical diffusivity is to
distribute the error over the entire box. We do this by shifting the data cube
by one grid point in the radial direction at the beginning of each (main)
time-step. In the time interval $\Delta t = N_x \delta t$ a physical structure
will have experienced numerical diffusivity at all radial locations in the box,
and since this is much shorter than the relevant physical time-scale
$2\pi\varOmega^{-1}$, the total numerical diffusivity should be approximately
the same for all flow features. In the third plot of \Fig{f:xdependent_error}
we show how all grid points have the same amplitude error, even after a long
integration time. The time-step may furthermore be lowered artificially in
order to reduce the time spent at any given radial location. This method is
preferred over the Random Revolution Center scheme, but at the cost of a
slightly higher numerical diffusivity as the shearing radial boundary
conditions must be set twice each time-step. Also postprocessing and on-the-fly
diagnostic output is complicated by the rapid radial displacement of the grid.

\subsubsection{Shear Advection by Fourier Interpolation (SAFI)}

A third possibility is to reduce amplitude error in the advection scheme
through time integration of the Keplerian advection term by interpolation. The
Shear Advection by Fourier Interpolation scheme is described in detail in
\App{s:safi} where we focused on making to allowed time-step longer by
integrating the Keplerian advection term through high order interpolation. This
scheme however also reduces the amplitude error of the advection to zero,
yielding a numerical diffusivity that does not depend on the radial coordinate.
The last panel of \Fig{f:xdependent_error} shows that the amplitude error is
indeed zero with this scheme.

In fact the SAFI method advects all scales with zero amplitude error and zero
dispersion error. The only exception is the Nyquist scale, which is damped as
the wave approaches a cosine function of $y$. This process of damping power at
the Nyquist scale can have a stochastic element to it, as it may take
arbitrarily long for the sine wave to be mapped as a cosine wave on the grid.

So in effect there will be an $x$-dependent amplitude error at the Nyquist
scale. This is however unlikely to affect the overall diffusivity of the
Keplerian advection, since the Nyquist scale is already heavily damped by the
explicit (hyper, shock, or regular) diffusivity terms. A combination of the SRD
and SAFI schemes would in theory remove the $x$-dependent damping of the
Nyquist scale, but we have not explored this any further since the SRD scheme
complicates postprocessing significantly.

\subsection{Comparison of Results Obtained with Various Shearing Box Schemes}

\begin{figure*}
   \includegraphics[width=\linewidth]{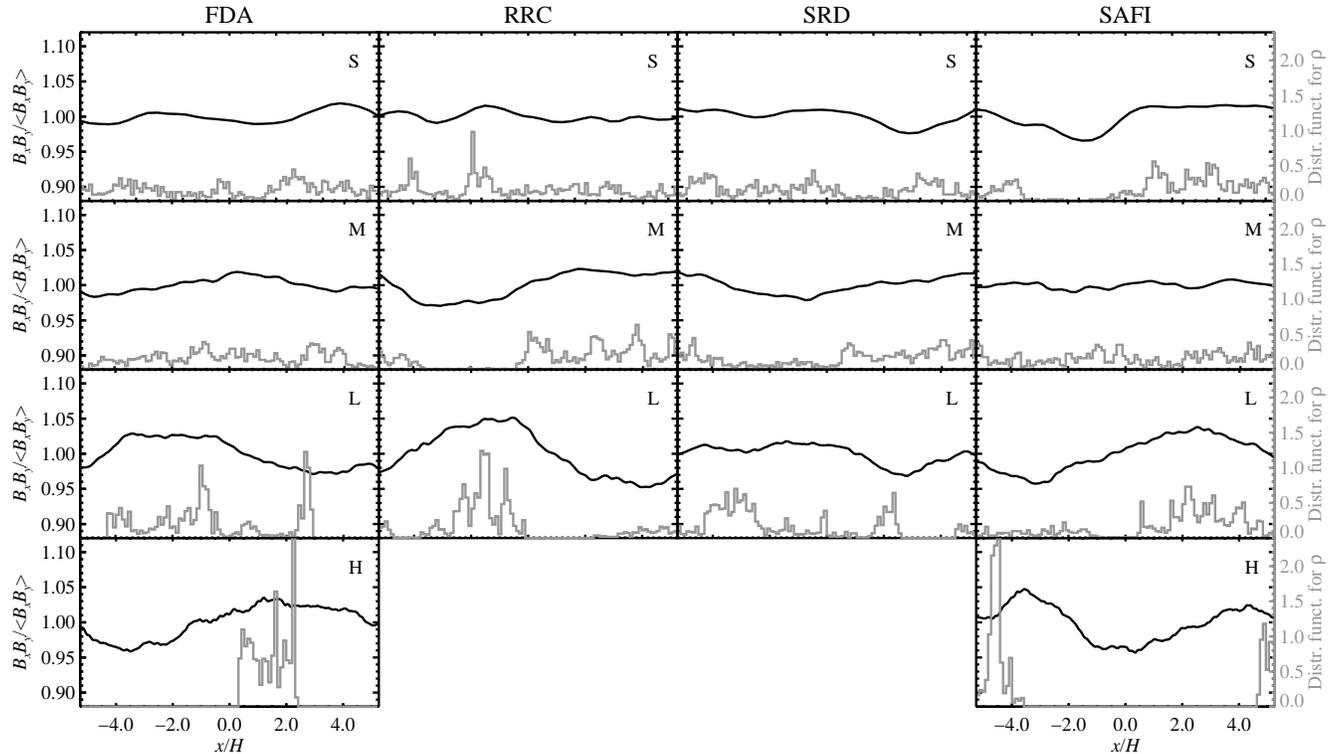}
   \caption{The Maxwell stress, divided by the mean Maxwell stress, as a
   function of the radial distance from the center of the box (black curve).
   Overplotted is the distribution function for the location of the minimum of
   the largest radial density mode (gray histograms). Rows represent various
   box sizes, while columns show the results for Finite Difference Advection
   (FDA), Random Revolution Center (RRC), Systematic Radial Displacement (SRD),
   and Shear Advection by Fourier Interpolation (SAFI). The typical fluctuation
   in the Maxwell stress is in all cases around 10\% (see \Tab{t:zonalflow}).
   In the smallest boxes these fluctuations have almost canceled out each
   other, since the correlation time of the stress is so short. The two largest
   boxes, on the other hand, have an excess of a few percent. Here the 100
   orbits integration has not covered enough coherence time-scales to give a
   good average. There is a slight tendency for the turbulence to be weak near
   the edges of the box, and for the density minimum to appear near the center
   of the box, when using the standard finite difference advection scheme (FDA,
   first column).}
  \label{f:bxbymx}
\end{figure*}
To test the relevance of space-dependent numerical dissipation for our results
we have run four variations of each of the standard box sizes S, M, L (FDA,
RRC, SRD, SAFI), and two variations of the largest box H (FDA, SAFI). The
Maxwell stress, normalized by the mean Maxwell stress, is shown in
\Fig{f:bxbymx} as a function of radial distance from the center of the box. We
have overplotted the distribution function for the location of the minimum of
the largest radial density mode (see histograms and the right axis). For the
standard finite difference advection scheme there is a tendency for the
turbulence to be weaker near the edges of the box and for the density minimum
to occur near the center of the box (see first column). However, the preference
for turbulence to be strong near the box center is not consistently present
with any of the improved shearing box algorithms. Thus we conclude that
space-dependent numerical diffusivity can potentially lead to spurious flow
features for large shearing boxes with the Pencil Code, but that this problem
is alleviated by either displacing the grid systematically in the radial
direction at each time-step or by integrating the Keplerian advection term by
Fourier interpolation.

The measured statistical properties of zonal flows, pressure bumps and large
scale power for various shearing box algorithms are written in
\Tab{t:zonalflow2}. For each box size there is a tendency for the large scale
power in zonal flows and pressure bumps to decrease when going from the
standard finite difference advection scheme to the Fourier interpolation
method. The amplitude of the Maxwell stress, on the other hand, is relatively
independent of the chosen algorithm. Based on these measurements we found it
safer to use the Fourier interpolation scheme for box sizes L and H.

\end{document}